\documentclass{article}

\usepackage{arxiv}

\usepackage[utf8]{inputenc} 
\usepackage[T1]{fontenc}    
\usepackage{hyperref}       
\usepackage{url}            
\usepackage{booktabs}       
\usepackage{amsfonts}       
\usepackage{nicefrac}       
\usepackage{microtype}      
\usepackage{lipsum}		
\usepackage{graphicx}
\usepackage{natbib}
\usepackage{doi}
\usepackage{amsmath}
\usepackage{caption}
\usepackage{subcaption}
\usepackage{comment}
\usepackage{array}
\usepackage{multirow}
\usepackage{appendix}

\title{Novel Models for Multiple Dependent Heteroskedastic Time Series}

\author{ \href{https://orcid.org/0009-0003-0427-368X}{\includegraphics[scale=0.06]{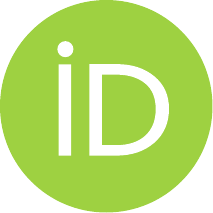}\hspace{1mm}Fangyijie Wang} \\
	School of Mathematics and Statistics\\
	University College Dublin\\
	\texttt{fangyijie.wang@ucdconnect.ie} \\
	\And
	\href{https://orcid.org/0000-0001-6232-9109}{\includegraphics[scale=0.06]{orcid.pdf}\hspace{1mm}Michael Salter-Townshend} \\
	School of Mathematics and Statistics\\
	University College Dublin\\
	\texttt{michael.salter-townshend@ucd.ie} \\
}

\renewcommand{\headeright}{}

\newcommand{\RNum}[1]{\expandafter\MakeUppercase{\romannumeral #1\relax}}

\hypersetup{
pdftitle={A template for the arxiv style},
pdfsubject={q-bio.NC, q-bio.QM},
pdfauthor={David S.~Hippocampus, Elias D.~Striatum},
pdfkeywords={First keyword, Second keyword, More},
}

\begin{document}
\maketitle

\begin{abstract}
Functional magnetic resonance imaging or functional MRI (fMRI) is a very popular tool used for differing brain regions by measuring brain activity. It is affected by physiological noise, such as head and brain movement in the scanner from breathing, heart beats, or the subject fidgeting. The purpose of this paper is to propose a novel approach to handling fMRI data for infants with high volatility caused by sudden head movements. Another purpose is to evaluate the volatility modelling performance of multiple dependent fMRI time series data. The models examined in this paper are AR and GARCH and the modelling performance is evaluated by several statistical performance measures. The conclusions of this paper are that multiple dependent fMRI series data can be fitted with AR + GARCH model if the multiple fMRI data have many sudden head movements. The GARCH model can capture the shared volatility clustering caused by head movements across brain regions. However, the multiple fMRI data without many head movements have fitted AR + GARCH model with different performance. The conclusions are supported by statistical tests and measures. This paper highlights the difference between the proposed approach from traditional approaches when estimating model parameters and modelling conditional variances on multiple dependent time series. In the future, the proposed approach can be applied to other research fields, such as financial economics, and signal processing. Code is available at \url{https://github.com/13204942/STAT40710}.
\end{abstract}


\section{Introduction}
In the statistical analysis of time series, the ARMA model describes stationary stochastic process with two polynomials, one is autoregression (AR), and the other one is moving average (MA) \cite{Moran:2018}. The GARCH model describes the conditional variance of the current error term related to the squares of previous innovations. Since such a model is used to model time series that exhibit volatility clustering across time \cite{Corbyn:2011} and forecast variances across time, the family of GARCH models has been widely developed \cite{Tim:1986}.

In recent years, many researchers have achieved outstanding performance in applications of ARMA and GARCH models in different research fields, e.g., forecasting stock index by applying ARIMA + GARCH models in 2021 \cite{Zolfaghari:2021}; obtaining corresponding fluctuation characteristics and forecasting transport flow by ARIMA + GARCH models \cite{Lin:2021}; and so on.

In neural science, researchers are interested in differing intrinsic timescales of brain regions. Takuya Ito used the autocorrelation (AR) function to perform the analysis \cite{Ito:2020}. John Fallon characterised BOLD signal dynamics by comparing over $6,000$ neural activity time series data. Typically, the fMRI data in different brain regions are modelled with AR($u$) or an ARMA($1,1$) process \cite{Purdon:2001}. However, the BOLD signal variability is sensitive to age differences and cognitive function \cite{Grady:2014}. In terms of analysing infants’ brain fMRI data, similar technique approaches are still applied. Certainly, some challenges cannot be ignored as they are specific to infants’ brain fMRI studies. Thus, more novel technical approaches are required to be proposed. One obvious challenge is motion. In general, head movement presents a substantial challenge to fMRI research. Infants tend to squirm and hard to stay still, which results in that head movement becoming problematic in infants. As a result, the infants’ brain fMRI data shows a presence of sudden timescale volatility. Nevertheless, few studies have directly explored the relationship between timescale volatility in infants’ different brain regions. Therefore, this research project focuses on differing and modelling the dependent volatility clustering across different brain regions by examining heteroscedasticity arising from head motion.

Many researchers have performed statistical approaches and model ARMA + GARCH for analysing time-varying variances and correlations primarily in financial time series analysis. This is because economic time series has the unique features of volatility clustering and limitations of using ARMA models. The novel aspect of this research project is adapting such approaches to differing volatility in fMRI data for infants.

This research project is concerned with the statistical analysis of functional magnetic resonance imaging (fMRI) time series data for infants. Functional MRI is widely used to measure brain activity. The technology behind it relies on the theory that blood flows increase in the area of the human brain which is in use \cite{Logothetis:2001}. A common method to perform fMRI data is blood oxygenation level- dependent (BOLD). The reason is that neural activity triggers changes in brain blood volume, flow, and oxygenation. However, this mechanism is not fully understood \cite{Logothetis:2002}. The fMRI data can be coded as a set of time series. Each time series contains observed BOLD signal variation for a single voxel (a unit representing the signal in brain scans on a three-dimensional grid) with the same time points acquired in a single session. A single voxel represents a tidy cube of brain tissue that can consist of a million brain cells. The researchers can perform statistical analysis of fMRI data to determine the signals induced by neural activity and noise. The BOLD signal observed in the measured signal is used to infer task-related activations in specific brain regions. Many studies have successfully captured task- evoked neural response patterns identified as associated with cognitive processes in brain regions \cite{Kanwisher:1997,Haxby:2001}.

Because the BOLD signal in the brain is sensitive to fluctuations non-neuronal activity, the head motion has significant, systematic effects on fMRI network measures. Different levels of head motion cause a difference in BOLD signal that could be considered neuronal activity by mistake \cite{VanDijk:2012}. Typically, the fMRI data can be modelled by using the time series Autoregressive model AR ($u$) or Autoregressive–moving-average model ARMA ($u,v$). Note that infants tend to move more than adults when they have fMRI scans for the brain, resulting in sudden volatility of infants’ fMRI data over time. However, few studies perform statistical analysis on this type of time series data.

This research studies modelling problems for a cluster of associated time series data sets with shared volatility over time. The proposed novel statistical approach fits the model by infant’s fMRI time series data of voxels. The research is challenging as it aims to build up multiple complex models constructed of independent AR($u$) models and one shared Generalized Autoregressive Conditional Heteroskedasticity GARCH ($p,q$) model. Equivalently, the white noise term of each AR ($u$) model is modelled by the same GARCH ($p,q$) model. No published studies show abilities to successfully identify the correct orders of AR + GARCH and estimate the model parameters. Therefore, the main concern in this research includes model identification, model diagnostics and parameter estimation for the multiple associated models ARMA ($u,v$) + GARCH ($p,q$).

Many Neural Sciences researchers will benefit from this research since they have shown high interest in measuring the activity of different brain regions by exploring fMRI data \cite{Ito:2020}. Moreover, the novel approach is extended to modelling finance time series data with sudden volatility caused by exogenous shocks, e.g., company share value. The researchers can adapt this novel approach to perform statistical analysis of complex nonlinear data with volatility in various research fields.

\section{Theory}
\subsection{ARMA time series models}
If a series is partly autoregressive and partly moving average, then it is a general time series model. It can be named a mixed autoregressive moving average model of orders $p$ and $q$, ARMA($p,q$). It is generally represented as below equation \cite{Corbyn:2011}:
\begin{equation}\label{eq:1}
Y_t=\phi_1 Y_{t-1}+\phi_2 Y_{t-2}+\cdots+\phi_p Y_{t-p}+e_t-\theta_1 e_{t-1}-\theta_2 e_{t-2}-\cdots-\theta_q e_{t-q}
\end{equation}

For the general ARMA($p,q$) model, the condition of stationary is $|\phi| < 1$ \cite{Corbyn:2011}. To identify the value of p and q, the sample autocorrelation function (ACF) in Equation \ref{eq:2.1} and the partial autocorrelation function (PACF) in Equation \ref{eq:2.2} is applied \cite{Corbyn:2011}:
\begin{equation}\label{eq:2.1}
\rho_k=\frac{\sum_{t=1}^{n-k}\left(Y_t-\bar{Y}\right)\left(Y_{t+k}-\bar{Y}\right)}{\sum_{t=1}^n\left(Y_t-\bar{Y}\right)^2} \quad k=1,2, \cdots
\end{equation}

\begin{equation}\label{eq:2.2}
\phi_{k, k}=\frac{\rho_k-\sum_{j=1}^{k-1} \phi_{k-1, j} \rho_{k-j}}{1-\sum_{j=1}^{k-1} \phi_{k-1, j} \rho_j} \quad j=1,2, \cdots, k-1
\end{equation}

For ARMA($p,q$), to identify $p$, Quenouille (1949) showed that the following approximation holds \cite{Corbyn:2011} for a white noise process:
\begin{equation}\label{eq:2.3}
    \operatorname{var}\left(\phi_{k, k}\right) \approx \frac{1}{n}
\end{equation}
Thus $\pm2/n$ is the critical limit on $\phi_{k,k}$ to test the null hypothesis that an AR($p$) model is correct. If $\phi_{k,k}$ breaks down to 0 for some $k$, there is evidence for $p = k$. To identify $q$, $\pm2/n$ can also be used as the critical limits on $\rho_k$ . The null hypothesis can be rejected if and only if $\rho_k$ exceeds these limits. And it shows evidence for $q = k$.

The sample ACF and PACF are practical visual tools for identifying orders of AR and MA models. But for a mixed ARMA model, the ACF and PACF have infinitely nonzero values. The summary of ACF and PACF behaviours are shown in Table \ref{tab:sum_acf_pacf}. So it is challenging to identify $p$ and $q$ \cite{Corbyn:2011}. Although other graphical tools are proposed to support us in identifying p and q, in this project, only AR($p$) model or MA($q$) model is considered instead of a mixed ARMA($p,q$) model. This is due to the invertibility characteristic of the AR and MA model, which is explained in section \ref{Invertibility}.

\begin{table}
    \centering
    \begin{tabular}{llll}
        \toprule
        & AR($p$) & MA($a$) & ARMA$(p,q), p>0$, and $q>0$\\
        \midrule
        ACF & Tails off & Null after lag $q$ & Tails off\\
        PACF & Null after lag $p$ & Tails off & Tails off\\
        \bottomrule
    \end{tabular}
    \caption{The summary of ACF and PACF behaviours for different time series models}
    \label{tab:sum_acf_pacf}
\end{table}

\subsection{Invertibility} \label{Invertibility}
An MA ($q$) can be invertible, which means it can be inverted into an infinite-order AR model. For instance, an MA (1) model is considered as:
\begin{equation}\label{eq:2.4}
Y_t=\varepsilon_t-\theta \varepsilon_{t-1}
\end{equation}

Equation \ref{eq:2.4} can be rewritten as:
\begin{equation}
\varepsilon_t=Y_t+\theta Y_{t-1}+\theta^2 Y_{t-2}+\cdots
\end{equation}
or
\begin{equation}
Y_t=\left(-\theta Y_{t-1}-\theta^2 Y_{t-2}-\theta^3 Y_{t-3}-\cdots\right)+\varepsilon_t
\end{equation}
by continuously replacing $t$ with $t - 1$ and substituting for $\epsilon_{t-1}$ when $|theta| < 1$ \cite{Corbyn:2011}. For a general MA ($q$) or ARMA($p,q$) model, if it is invertible, it can be inverted to an AR($p$) model with a large p. In this project, the proposed model will only have AR($p$) component instead of ARMA($p,q$). When $p$ is increased to a large value, the model equivalently has an ARMA($p,q$) component.

\subsection{Time series model of heteroscedasticity} \label{heteroscedasticity}
Consider a single time series data $Y_t$ with high volatility, the conditional variance of $Y_t$ is given by the past $Y$ values, $Y_{t-1}, Y_{t-2}, \cdots$. In practice, the one-step-ahead conditional variance varies with the current and past values. Therefore, the conditional variance is a random process. To study the volatility of a time series, applying the McLeod-Li test Field \cite{McLeod:1983} for the presence of volatility is useful. In 1982, Engle first proposed the autoregressive conditional heteroscedasticity (ARCH) model for modelling the changing variance of a time series \cite{Engle:1982}. The null hypothesis of the McLeod- Li test is that no autoregressive conditional heteroskedasticity (ARCH) is present among the lags considered. It is computing the squared series data or the squared residuals from an ARMA model and then performing Ljung-Box test \cite{LJUNG:1978} with the calculated results. When there is $k$ out of $n$ p-values significant at the $0.05$ significance level among lags $n$, and $\frac{k}{n} > 0.05$, the null hypothesis is likely to reject.

Another popular model to represent the dynamic evolution of volatility in time series is the Generalized Autoregressive Conditional Heteroscedasticity (GARCH) model \cite{Tim:1986}.
\begin{equation} \label{eq:2.3.3}
\begin{split}
Y_t &= \sigma_{t \mid t-1} \varepsilon_t \\
\sigma_{t \mid t-1}^2 &= \alpha_0+\alpha_1 Y_{t-1}^2+\cdots+\alpha_q Y_{t-q}^2+\beta_1 \sigma_{t-1 \mid t-2}^2+\cdots+\beta_p \sigma_{t-p \mid t-p-1}^2 \\
\varepsilon_t &\sim N(0,1) \\
\end{split}
\end{equation}

The orders of GARCH are $p$ and $q$. In Equation \ref{eq:2.3.3}, the standardized residuals $\hat{\varepsilon}_t$ are computed as $Y_t / \sigma_{t \mid t-1}$. If the GARCH model is correct, $\hat{\varepsilon}_t$ is independent and identically distributed \cite{Corbyn:2011}. To identify $p$ and $q$ in GARCH($p,q$), the method examines the ACF and PACF of squared ${Y_t}$. If ${Y_t}$ is following GARCH($p,q$), then ${Y^2_t}$ is following ARMA($max(p,q),p$) \cite{Corbyn:2011}. However, sometimes it is difficult to identify p and q because of more fluctuation and high variance in data. Some existing well-known Information Criteria, Akaike Information Criteria (AIC), Bayesian Information Criteria (BIC), can help us choose the correct model from a list of candidate GARCH($p,q$) models \cite{Naik:2020}. The lower AIC or BIC, the candidate GARCH model is better. The AIC value can be calculated from the maximum likelihood estimate of the GARCH model, and it is defined as:
\begin{equation} \label{eq:2.3.4}
AIC = 2K - 2\ln(L)
\end{equation}
In Equation \ref{eq:2.3.4}, $K$ is the number of independent variables used, and $L$ is the log-likelihood estimate of the candidate GARCH model.

\section{Methods}
This research project will implement the proposed statistical analysis approach in R programming language using the software RStudio.

Initially, a single time series data will be simulated from a model AR($1$) + GARCH($1,1$) with fixed orders and model parameters. The simulation time series data will be used for statistical analysis afterwards. The sample Autocorrelation Function (ACF) and the sample Partial Autocorrelation Function (PACF) (Corbyn, 2011) plots provide practical graphical tools for identifying the orders of AR($u$) or MA($v$) models. Fit the identified AR ($u$) model to the simulation data, and the squared residuals from this model can be used to test for the presence of ARCH (or GARCH). The McLeod-Li (McLeod and Li, 1983) test can refer to conditional heteroscedasticity (ARCH) effects by using several lags and plotting the p-value of the statistical tests. The GARCH model is an extension of the ARCH model that incorporates a moving average component together with the autoregressive part. The model identification techniques for ARMA models can also be applied to the squared residuals. Plotting ACF and PACF of the squared residuals typically indicates ARMA($max(p,q),p$) model is the suitable model for the squared residuals \cite{Corbyn:2011}. Thus, a GARCH($p,p$) model is fitted at first, and then $q$ can be estimated by examining the significance of the resulting ARCH coefficient estimates. After this, a model AR($u$) + GARCH($p,q$) is fitted to the simulation data to estimate the model parameters $\phi$, $\alpha$, $\beta$.

Similarly, the technical approach is applied to multiple dependent heteroskedastic time series. It starts with simulating data from various models with shared volatility clustering across time $y_i$ = AR($u_i$) + GARCH($p,q$) as below:

\begin{equation}
\begin{split}
    y_1&=\operatorname{AR}\left(u_1\right)+\operatorname{GARCH}(p, q) \\ 
    y_2&=\operatorname{AR}\left(u_2\right)+\operatorname{GARCH}(p, q) \\ y_3&=\operatorname{AR}\left(u_3\right)+\operatorname{GARCH}(p, q) \\
    &\cdots \cdots
\end{split}
\end{equation}

Then sample ACF and sample PACF are used to identify each AR orders $u_i$. A collection of AR($u_i$) models are fitted to each time series separately. Therefore, a series of estimated residuals $\hat{\eta}_i$ returned from each time series $\{y_i\}$ is obtained. The next step is averaging over all the $N$ series ($\frac{1}{N}\sum_{i=1}^N \hat{\eta}_i$) to obtain the average value of all estimated residuals. Note that the estimated residuals are shared components in all $N$ series. Calculate the average of $N$ estimated residuals $\{{\hat{\eta}_t}\}$ as $\bar{\hat{\eta}}_t$ and plot ACF and PACF of the $\bar{\hat{\eta}}_t$ so that the GARCH orders can be identified. Equivalently, the model identification techniques for ARMA models are used to identify $p$ and $max(p,q)$. When fitting the GARCH($p,q$) to the average value of $N$ estimated residuals $\{\bar{\hat{\eta}}_t\}$, the estimated coefficients $(\alpha_i, \beta_i)$ are obtained by maximising the likelihood function of the GARCH model \cite{Corbyn:2011}.
\begin{equation}
\mathrm{L}(\widehat{\omega}, \widehat{\alpha}, \widehat{\beta})=-\frac{n}{2} \log (2 \pi)-\frac{1}{2} \sum_{i=1}^n\left\{\log \left(\hat{\sigma}_{t \mid t-1}^2\right)+\hat{\eta}_t^2 /\left(2 \hat{\sigma}_{t \mid t-1}^2\right)\right\}
\end{equation}

There is no closed-form solution for the maximum likelihood estimators of $\widehat{\omega}$, $\widehat{\alpha}$ and $\widehat{\beta}$, but they can be computed numerically. A series of models can be built, including a shared GARCH model plus multiple independent AR($u$) models. Afterwards, fitting such a series of models to fMRI data performs the goodness of fit test for the fitted AR + GARCH model.

\subsection{Simulate data}
The simulation work starts with simple models and fixed parameters. The objectives are simulating multiple time series data sets with given parameters and orders. The simulated data will be used for model fitting and parameter estimation.

The mixed model AR($u$) + GARCH($p,q$) is proposed. In general, the model is used for modelling multiple dependent series data, so it contains multiple AR + GARCH formulas that can be written as:
\begin{equation}
\begin{split}
    Y_{1t} &= \mu + \sum_{i=1}^u \phi_{1i} Y_{t-i} + \eta_{1t} \\
    Y_{2t} &= \mu + \sum_{i=1}^u \phi_{2i} Y_{t-i} + \eta_{2t} \\
    &\vdots \\
    Y_{kt} &= \mu + \sum_{i=1}^u \phi_{ki} Y_{t-i} + \eta_{kt} \\
\end{split}
\end{equation}
And the squared residuals $\{\eta_{kt}\}$ are shared with each formula. It is assumed that the average of $\{\eta_{kt}\}$, $\bar{\eta}_t$, can be fitted by a GARCH($p,q$) model. Thus, its formula is written as:
\begin{equation}
\begin{split}
    \bar{\eta}_t &= \sigma_{t \mid t-1} \varepsilon_t \\
    \sigma_{t \mid t-1}^2 &= \alpha_0+\alpha_1 \eta_{t-1}^2+\cdots+\alpha_q \eta_{t-q}^2+\beta_1 \sigma_{t-1 \mid t-2}^2+\cdots+\beta_p \sigma_{t-p \mid t-p-1}^2 \\
\end{split}
\end{equation}

In the beginning, the orders \{$u, p, q$\} are given fixed values \{$1, 1, 1$\}. Therefore, the simple model is AR($1$) + GARCH($1,1$). The coefficient parameter $\varphi$ of AR model is given $0.05$, close to $0.0$. The coefficient parameter \{$\alpha, \beta$\} of GARCH model are randomly given fixed values \{$0.2, 0.5$\}. In total, 20, 100 and 400 time series data are simulated with this AR($1$) + GARCH($1,1$) model. These time series data are used to assess whether the size of the data set impacts the results of parameter estimation.

Afterwards, another AR($1$) + GARCH($1,1$) model is created with the parameters as the previous simulation, except parameter $\varphi$. The parameter $\varphi$ of AR model is set as a random number in a range ($0.7, 0.9$), close to $1.0$. This new model is used for simulating 400 time series data sets. Parameter estimation is performed on these data sets. Then the results are compared with previous parameter estimation with 400 time series data with fixed $\varphi$. The objective of this simulation work is to assess whether parameter estimation results behave differently when the parameter $\varphi$ is a fixed value and a random value.

In the end, 400 time series data are simulated with AR($1$) + GARCH($1,1$) model. But the first 200 time series have $\alpha = 0.2, \beta = 0.5$, and $\varphi$ is a random number in range ($0.01, 0.05$). The reset 200 time series have the same \{$\alpha, \beta$\} as \{$0.2, 0.5$\}. The only difference is that parameter $\varphi$ is in the range ($0.7, 0.9$). These 400 time series data together are used for parameter estimation. The objective is to understand the behaviour of parameter estimation when the parameter $\varphi$ is close to 0.0 and 1.0.

All simulated time series data contain 300 time points ($N = 300$) as default.

\subsection{Model identification and parameter estimation}
The following estimation work is programmed in R language using RStudio.

When 20, 100, 400 time series data are simulated, each time series’ ACF and PACF values are calculated in a created loop function. (1) Comparing the ACF and PACF among the first 20 lags to count the number of PACF values which are significantly not null ($\pm2/n$ is the critical limit) when ACF decays exponentially fast. The number of these non-null PACF values $k$ is equal to the AR orders, $p$. The estimated model is named AR ($\hat{p}$). If only PACF decays fast, then counting the number of ACF values that are not equal to null. This number $k$ is equal to the MA orders, $q$, so the estimated model is named MA($q$). If ACF and PACF decay fast and exponentially, this is evidence that the ARMA model is present. But the fixed model ARMA is not considered at the stage of this project. (2) After the estimated AR model is fitted on each series $Y_{it}$ , a set of squared residuals \{$\hat{\eta}_{it}$\} is returned. Then, compute the average of \{$\hat{\eta}_{it}$\} for 20, 100 and 400. (3) Afterwards, the ACF and PACF of the averaged $\bar{\hat{\eta}}_t$ are calculated for identifying GARCH orders $p$ and $q$. It is the same identification method used for identifying AR orders. The estimated GARCH($p,q$) model is fitted on averaged $\bar{\hat{\eta}}_t$ and model diagnostic is performed afterwards. The goal of the diagnostic is to measure the goodness of fit of the estimated GARCH model. (4) Each series data $Y_{it}$ minus averaged $\bar{\hat{\eta}}_t$ before calculating ACF and PACF of series again. This step is to identify the AR orders $p$ again and estimate parameter $\varphi_i$ after each series $Y_{it}$ is fitted by the estimated AR($p$) model.

The previous four analysis steps are also executed on the 400 simulated time series data with parameter $\varphi_i$ in range ($0.7, 0.9$). (1) Computing ACF and PACF of each series $Y_{it}$ and checking their behaviour to identify the order of the AR model for $Y_{it}$ ($\pm2/N$ is the critical limit). (2) Once the AR model is fitted on each series, the returned residuals \{$\hat{\eta}_{it}$\} from each fitted model are collected. (3) The averaged residuals $\bar{\hat{\eta}}_t$ is computed from the collection of returned residuals \{$\hat{\eta}_{it}$\}. After this, the $\bar{\hat{\eta}}_t$ is removed from each series $Y_{it}$ and ACF and PACF of $\bar{\hat{\eta}}_t$ are calculated for identifying orders of GARCH model $p$ and $q$. The estimated GARCH($p,q$) model is fitted on $\bar{\hat{\eta}}_t$ before performing model diagnostic. (4) Calculating ACF and PACF on each series data $Y_{it}$ to identify orders of the AR model $u_i$. To estimate coefficient parameters $\varphi_i$, it can be retrieved from results of fitting the estimated AR($u_i$) model.

The methods used for analysing the mixed 400 simulated time series with variant parameters differ from previous methods. The reason is that some backwards are found when estimating parameters. The details of backwards will be explained in the following few sections. This new analysis methods are described as: (1) The ACF and PACF of series data $Y_{it}$ are computed for the first 20 lags. And they are used for identifying AR orders $p$ and $q$. Then \{$Y_{it}$\} is fitted with the estimated ARMA($\hat{p},\hat{q}$) model to get model residuals \{$\hat{\eta}_{it}$\}. Also, the coefficient parameters $\varphi_i$ are estimated after model fitting. All estimated $\hat{\varphi}_i$ are stored in a numerical vector $\hat{\varphi}_{1i}$ that is used for final estimation later. (2) A weight
parameter $W_i$ is calculated by using each estimated $\hat{\varphi}_{1i}$, and it can be written as:
\begin{equation}
\begin{split}
    w_i &= \frac{1}{\hat{\varphi}_{1i}} \\
    W_i &= \frac{w_i}{\sum w_i}
\end{split}
\end{equation}
Afterwards, the averaged residuals $\bar{\hat{\eta}}_t$ is calculated by multiplying the weight $W_i$ with each model residuals \{$\hat{\eta}_{it}$\}:
\begin{equation}
\bar{\hat{\eta}}_t = \sum_{i=1}^K W_i \hat{\eta}_{it}
\end{equation}

(3) Calculating the ACF and PACF of averaged residuals $\bar{\hat{\eta}}_t$ is the method used for identifying the orders of GARCH model $p$ and $q$. An estimated GARCH model can be built with $\hat{p}$ and $\hat{q}$. The model is fitted on the averaged residuals $\bar{\hat{\eta}}_t$ so that model diagnostic can be performed. (4) The averaged residuals $\bar{\hat{\eta}}_t$ is removed from each series $Y_{it}$. The ACF and PACF of each series \{$Y_{it} - \bar{\hat{\eta}}_t$\} are identified to build the estimated ARMA($\hat{p}, \hat{q}$). The estimated ARMA models are fitted on each \{$Y_{it} - \bar{\hat{\eta}}_t$\} to get the estimated coefficient parameters $\hat{\varphi}_2$. All $\hat{\varphi}_2$ are stored in numerical vector $\hat{\varphi}_{2i}$. (5) Average two numerical vectors $\hat{\varphi}_{1i}$ and $\hat{\varphi}_{2i}$ to get the final estimation of parameter $\hat{\varphi}_{l}$.

\subsection{Fit model on real-world data}
\begin{figure}
    \centering
    \includegraphics[width=1\linewidth]{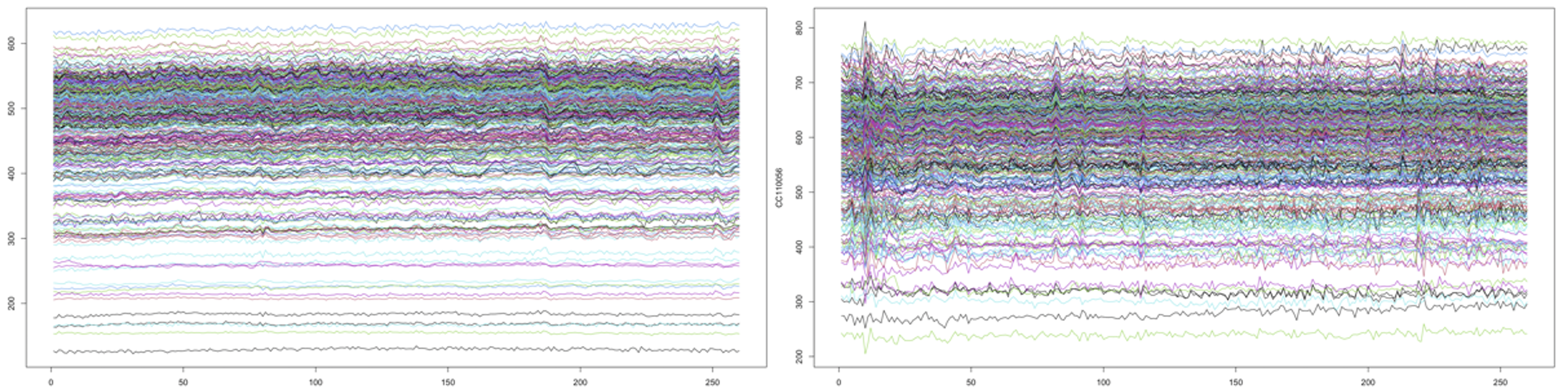}
    \caption{Subject CC110056 shows high volatility on the left. On the right, CC110045 fMRI shows very little movement.}
    \label{fig:real_data}
\end{figure}

The real-world data set is collected from two real fMRI data sets, subject CC110045 and subject CC110056. And subject CC110056 shows high volatility than subject CC110045. Each subject represents the time course for a different area, called Region of Interest (ROI) in the brain. There are 400 ROIs in each subject, and each ROI time series data contains 261 time courses. Typically, ROIs beside each other in the brain have high correlations. Both data sets have no missing values and no invalid numerical values, see Figure \ref{fig:real_data}. They are valid time series data examined by visualisation tools.

The improved analysis method is applied to real-world fMRI data. The estimation of parameters and model fitting are examined statistically. It includes estimated AR orders $\hat{p}_i$, estimated GARCH orders ($\hat{p},\hat{q}$), estimated AR coefficients $\hat{\varphi}_i$, the averaged residuals $\bar{\hat{\eta}}_t$, and standardized residuals $\hat{\varepsilon}_t$. Especially, to evaluate the new improved method, the coefficients estimator $\hat{\varphi}_i$ is compared to the old coefficients’ estimator $\hat{\varphi}_{i\_old}$ that are generated from the old statistical analysis method.

\section{Results and Discussion}
The idea of simulating multiple time series in different scenarios is to help us evaluate the performance of the proposed modelling methods and to understand the behaviour of numerous dependent time series data with volatility. In this section, there are some plots and tables to present the results of the modelling.

In the modelling, there are four critical parameters in consideration and comparison, $\hat{\eta}_t$, $\sigma_{t \mid t-1}$, $\varepsilon_t$ and $\varphi_i$. The first parameter $\bar{\hat{\eta}}_t$ is the averaged value of multiple residuals returned from AR models. It is possible to assess the modelling of $\bar{\eta}_t$ by evaluating the estimation of $\sigma_{t \mid t-1}$ and returned standardized residuals $\hat{\varepsilon}_t$, because Equation \ref{eq:2.3.3} shows that $\bar{\eta}_t$ is determined by the product of $\sigma_{t \mid t-1}$ and $\varepsilon_t$. As our assumptions state, the distribution of $\varepsilon_t$ follows a normal distribution. Thus, the distribution of $\hat{\varepsilon}_t$ is assessed by a normal quantile-quantile plot (Q-Q plot) \cite{Wilk:1968}. Comparing the estimated $\hat{\sigma}_{t \mid t-1}$ to the actual $\sigma_{t \mid t-1}$ to assess its accuracy of estimation. This method evaluates the estimation of $\sigma_{t \mid t-1}$ and examines the fitting performance of GARCH model.

Another critical parameter $\hat{\varphi}_i$ is estimated from the fitted AR model. Since the actual values of $\varphi_i$ are known, its estimation is assessed by computing the mean squared error (MSE) of $\hat{\varphi}_i$. And the scatter plot of $\hat{\varphi}_i$ and $\varphi_i$ is a visualisation tool to assess the estimation accuracy.

\subsection{Simulation study \RNum{1} - Different sizes of data sets}
When the parameters are given fixed values and the size of the simulated time series data set increases from 20 to 100, the estimation of $\sigma_{t \mid t-1}$ can be evaluated by the below plots:

\begin{figure}
     \centering
     \begin{subfigure}[b]{0.49\textwidth}
         \centering
         \includegraphics[width=\textwidth]{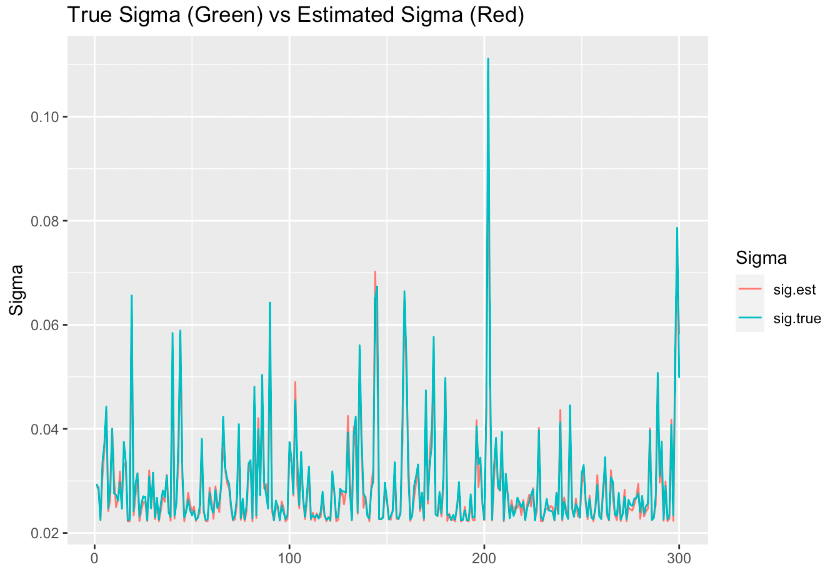}
         \caption{Plots of actual sigma $\sigma_{t \mid t-1}$ and estimated sigma $\hat{\sigma}_{t \mid t-1}$}
         \label{fig:essay_fig_41_1}
     \end{subfigure}
     \begin{subfigure}[b]{0.49\textwidth}
         \centering
         \includegraphics[width=\textwidth]{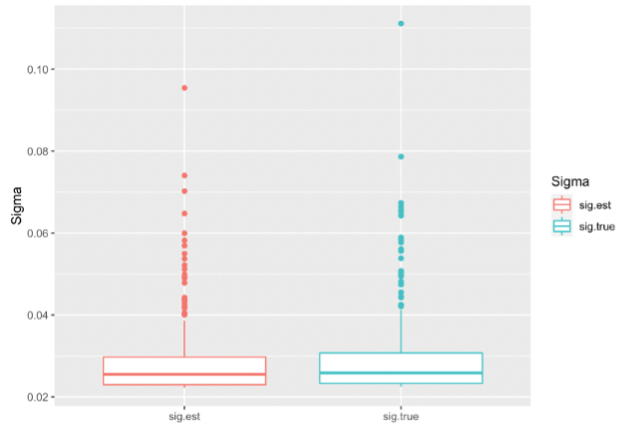}
         \caption{Boxplots of actual sigma $\sigma_{t \mid t-1}$ and estimated sigma $\hat{\sigma}_{t \mid t-1}$}
         \label{fig:essay_fig_42}
     \end{subfigure}
    \caption{}
    \label{fig:sigma_boxplots}
\end{figure}

The above plots show that the estimator $\hat{\sigma}_{t \mid t-1}$ are very close to the actual values of $\sigma_{t \mid t-1}$. And the mean value of the estimator $\hat{\sigma}_{t \mid t-1}$ is approximately equal to the expected mean of $\sigma_{t \mid t-1}$. Also, the plots indicate that the values of estimator $\hat{\sigma}_{t \mid t-1}$ do not change a lot when the size of the data set is increasing. When the size of data set increases to 400, the values of estimator $\hat{\sigma}_{t \mid t-1}$ are still close to the actual values of $\sigma_{t \mid t-1}$ without significant variance. Thus, the estimation of $\sigma_{t \mid t-1}$ is quite good.

When the size of data sets increases, the distribution of standard residuals $\hat{\varepsilon}_t$ follows a normal distribution. The Q-Q plots of estimator $\hat{\varepsilon}_t$ are shown in Figure \ref{fig:essay_fig_43}:
\begin{figure}
    \centering
    \includegraphics[width=0.6\linewidth]{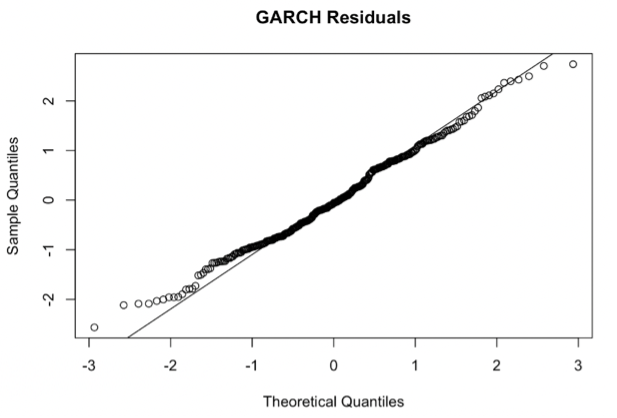}
    \caption{Q-Q plots of actual $\varepsilon_t$ and estimated residuals $\hat{\varepsilon}_t$}
    \label{fig:essay_fig_43}
\end{figure}

It also means that the assumption of the proposed model parameter $\hat{\varepsilon}_t$ is correct and accepted. Equivalently, the value of averaged estimator $\bar{\hat{\eta}}_t$ should be good as expected because it is only determined by values of $\hat{\sigma}_{t \mid t-1}$ and $\hat{\varepsilon}_t$.

\begin{figure}
     \centering
     \begin{subfigure}[b]{0.3\textwidth}
         \centering
         \includegraphics[width=\textwidth]{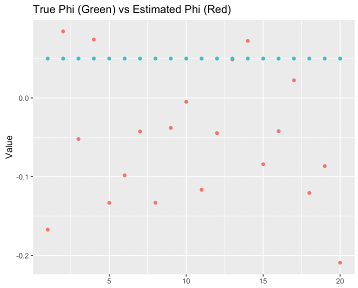}
         \caption{}
         \label{fig:table_41_1}
     \end{subfigure}
     \begin{subfigure}[b]{0.3\textwidth}
         \centering
         \includegraphics[width=\textwidth]{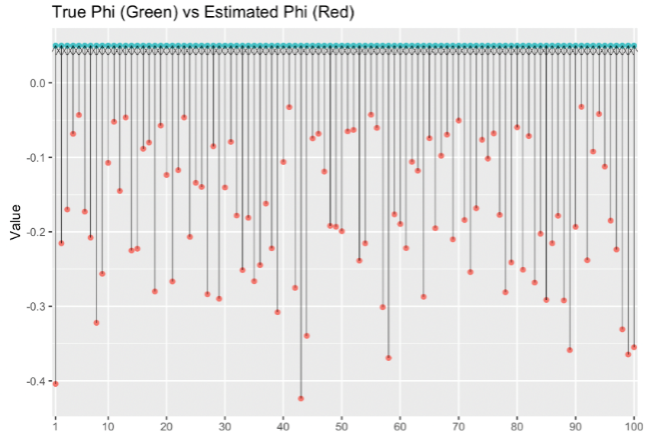}
         \caption{}
         \label{fig:table_41_2}
     \end{subfigure}
     \begin{subfigure}[b]{0.3\textwidth}
         \centering
         \includegraphics[width=\textwidth]{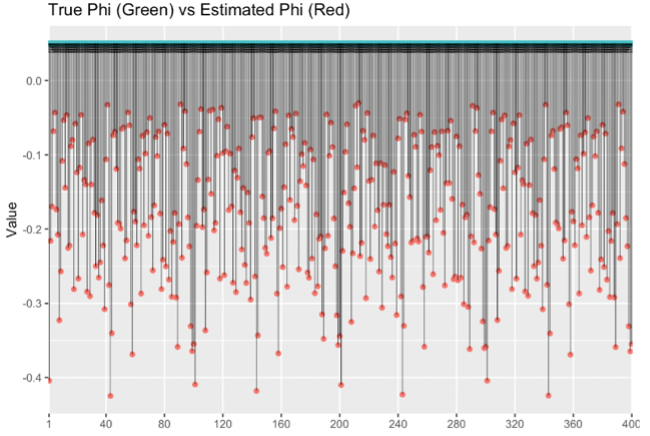}
         \caption{}
         \label{fig:table_41_3}
     \end{subfigure}   
    \caption{Comparison of true $\varphi_i$ to estimated $\hat{\varphi}_i$}
    \label{fig:phi_plots}
\end{figure}

Further, the orders of the averaged estimator $\bar{\hat{\eta}}_t$ is consistently identified as ($1,1$) when the size of data sets increases from 20 to 400. Table \ref{tab:table_41} presents that fitting the GARCH($1,1$) model on estimator $\bar{\hat{\eta}}_t$ always returns the lowest AIC using MLE. Although GARCH($2,2$) is fitted on $\bar{\hat{\eta}}_t$, the model GARCH($1,1$) achieves the lowest AIC. Thus, GARCH($1,1$) is considered the best-fit model.
\begin{table}
    \centering
    \begin{tabular}{lll}
        \toprule
        & GARCH($1,1$) & GARCH($2,2$) \\
        \midrule
        AIC & -0.7293573 & -0.7181245 \\
        \bottomrule
    \end{tabular}
    \caption{Comparison of AIC values between two models}
    \label{tab:table_41}
\end{table}
After removing $\bar{\hat{\eta}}_t$ from each series, the values of estimator $\hat{\varphi}_l$ varies differently. To compare the actual values to the estimated values, it is necessary to plot values with scatter plots, Figure \ref{fig:phi_plots}.

When looking at the first plot of $\varphi_i$ and estimator $\hat{\varphi}_l$, it is clear to see few estimated values larger than the actual value ($0.05$), and the reset of values are less than $0.05$. The estimator mainly underestimates the parameters $\varphi_i$. When the size of data sets increases to 400, all estimator values $\hat{\varphi}_l$ are less than the actual value (0.05). Referring to these plots, the estimation of $\varphi_i$ is biased when the size of data sets is large.

\subsection{Simulation study \RNum{2} - Modelling with dynamic parameters}
In this section of study \RNum{2}, some statistical analysis results are presented to compare modelling performance when $\varphi_i$ is fixed and dynamic.

To compare estimator $\hat{\sigma}_{t \mid t-1}$ when $\varphi_i$ is in different scenarios, it is necessary to plot actual values and estimated values together to study.

\begin{figure}
    \centering
    \includegraphics[width=0.8\linewidth]{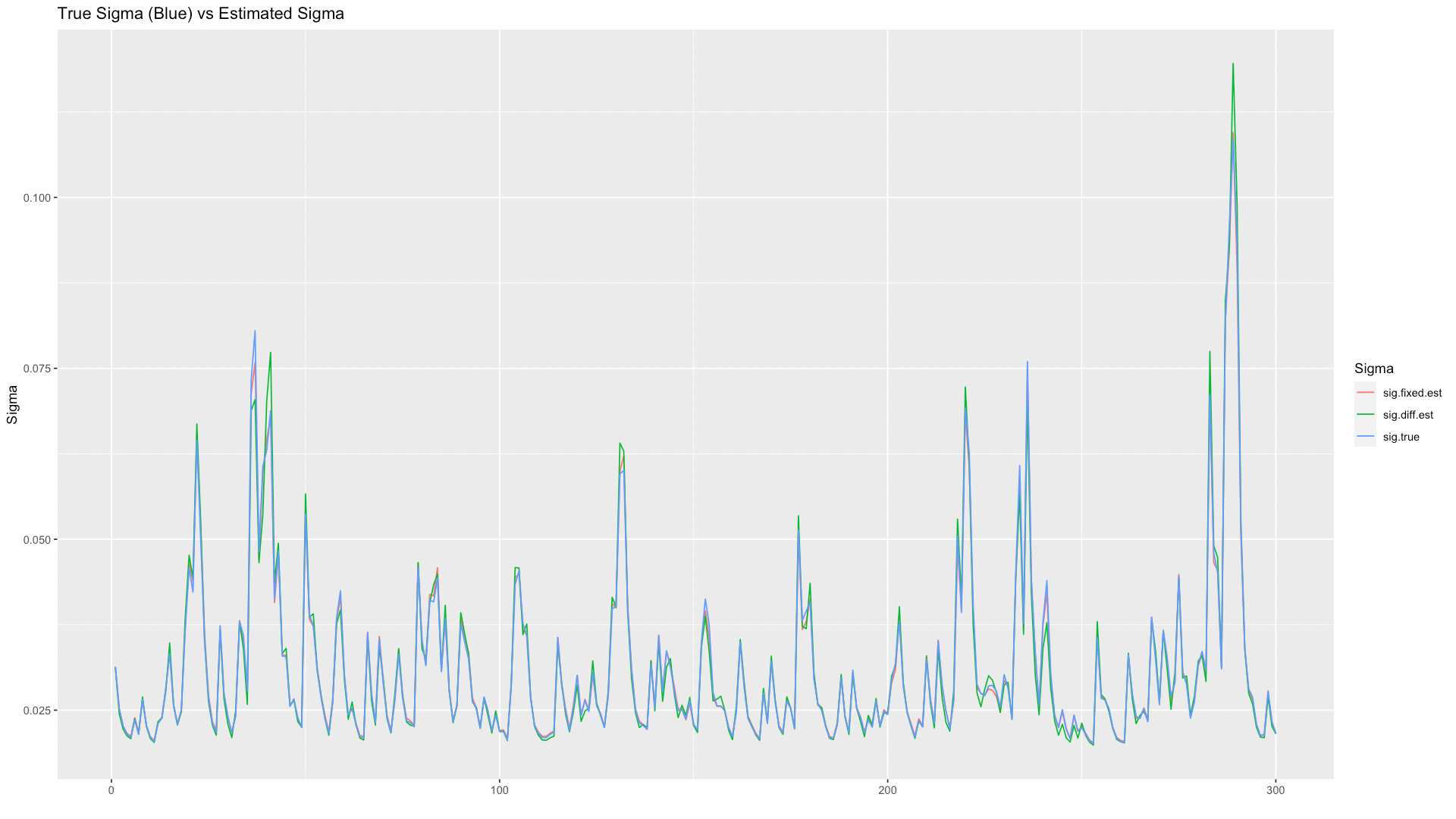}
    \caption{Plots of actual sigma $\sigma_{t \mid t-1}$ and estimated sigma $\hat{\sigma}_{t \mid t-1}$}
    \label{fig:essay_fig_45_1}
\end{figure}

Figure \ref{fig:essay_fig_45_1} demonstrates that the values of two different estimators $\hat{\sigma}_{t \mid t-1}$ are quite good, because both estimators fit the actual values. While the figure reveals some estimation errors on large $\hat{\sigma}_{t \mid t-1}$, the estimators do not perform well when estimating large values.

\begin{figure}
     \centering
     \begin{subfigure}[b]{0.49\textwidth}
         \centering
         \includegraphics[width=\textwidth]{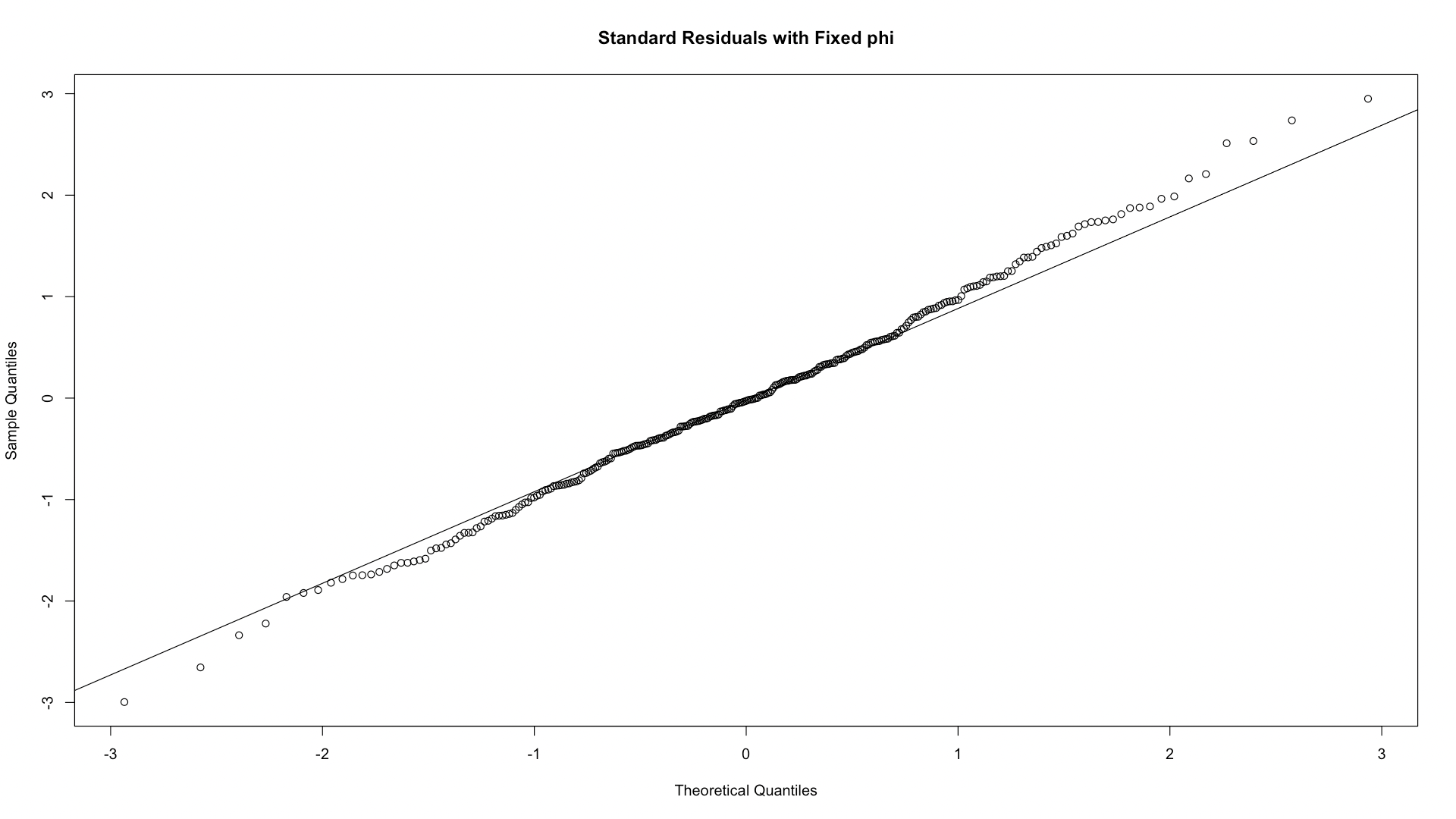}
         \caption{}
         \label{fig:essay_fig_46_1}
     \end{subfigure}
     \begin{subfigure}[b]{0.49\textwidth}
         \centering
         \includegraphics[width=\textwidth]{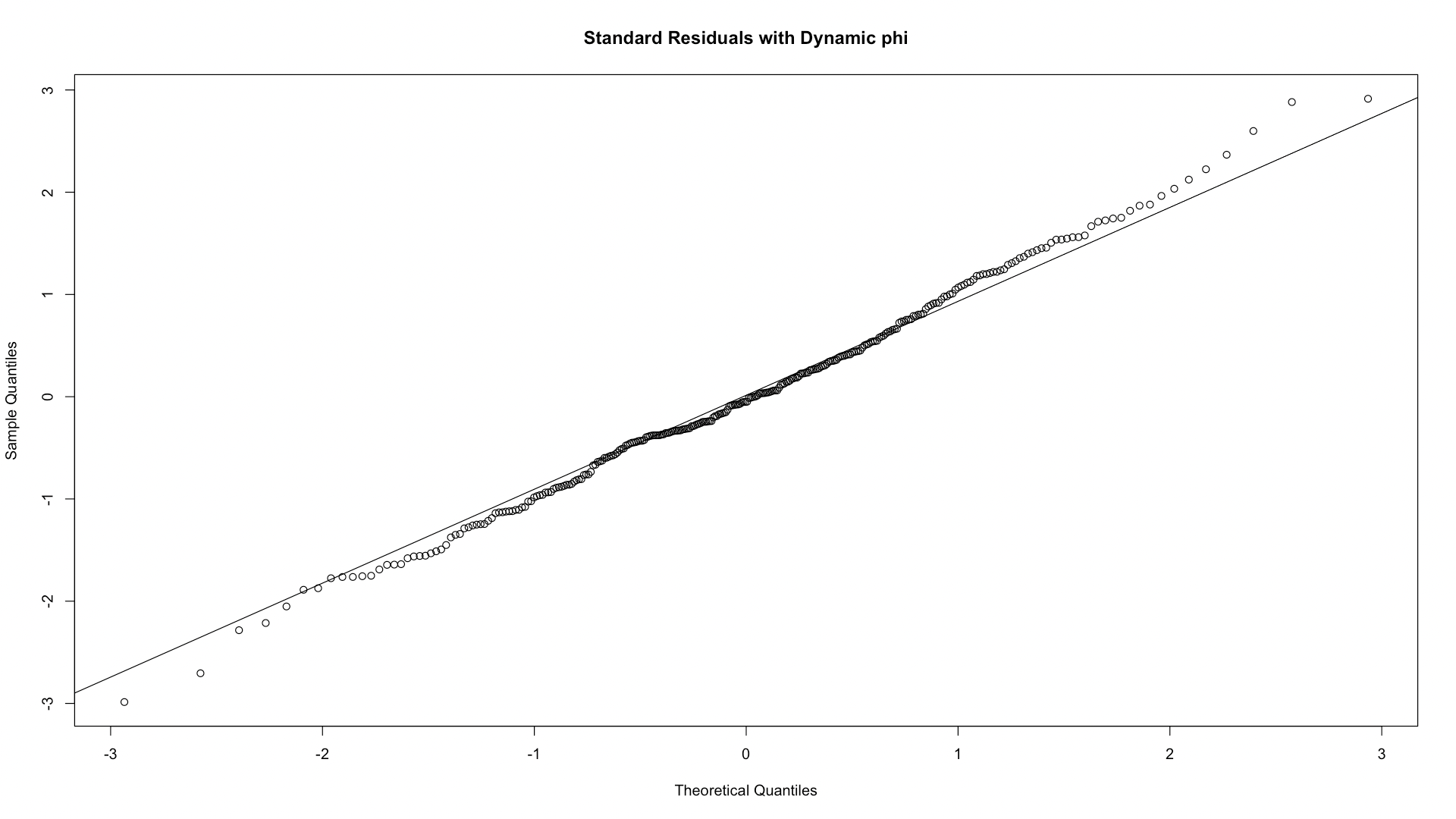}
         \caption{}
         \label{fig:essay_fig_46_2}
     \end{subfigure}
    \caption{plots of actual residuals $\varepsilon_t$ and estimated residuals $\hat{\varepsilon}_t$}
    \label{fig:varepsilon_plots}
\end{figure}

Similarly, the distribution of standardized residuals $\varepsilon_t$ can be examined by Q-Q plots. The above Q-Q plots, Figure \ref{fig:varepsilon_plots}, indicate that both estimators $\hat{\varepsilon}_t$ follow a normal distribution. The data points follow the central line very closely for estimators $\hat{\varepsilon}_t$.

Moreover, the orders of GARCH model are accurately identified as ($1,1$). AIC is used to compare GARCH($1,1$) with GARCH($2,2$) to prove the identification method works correctly.
\begin{table}
    \centering
    \begin{tabular}{lll}
        \toprule
        AIC & GARCH($1,1$) & GARCH($2,2$) \\
        \midrule
        $0.7 < \varphi < 0.09$ & -0.6760055 & -0.6653040 \\
        $\varphi=0.05$ & -0.6710866 & -0.6620987 \\
        \bottomrule
    \end{tabular}
    \caption{Comparison of AIC values between two models when $\varphi$ in different ranges}
    \label{tab:table_42}
\end{table}

According to AIC, the best-fit model explains the most significant amount of variation using the fewest possible independent variables. Regarding Table \ref{tab:table_42}, it gives apparent results that GARCH($1,1$) is always better than GARCH($2,2$) to fit shared volatility clustering $\hat{\varepsilon}_t$ no matters whether $\varphi_i$ is fixed to $0.05$ or dynamic in range ($0.7,0.9$). When only considering model GARCH($1,1$), it performs better when $\varphi_i$ is dynamic in range ($0.7,0.9$).

In terms of the estimator $\hat{\varphi}_l$, the results are presented by a scatter plot.

\begin{figure}
    \centering
    \includegraphics[width=0.8\linewidth]{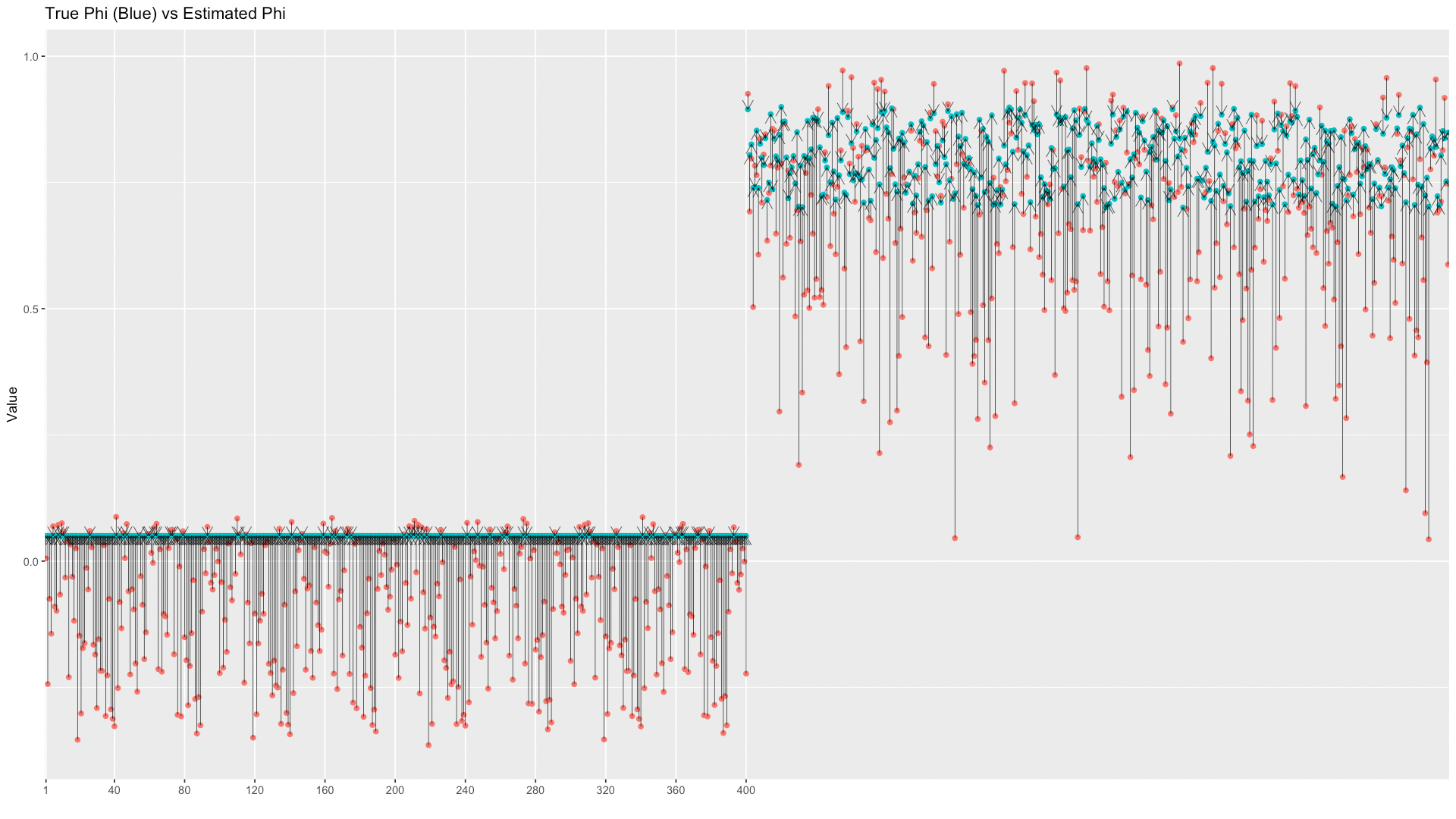}
    \caption{Comparison of true $\varphi_i$ to estimated $\hat{\varphi}_l$}
    \label{fig:essay_fig_47_1}
\end{figure}

For each $\varphi_i$, the scatter plot can illustrate the distance between the actual value and the estimated value. The scatter plot, Figure \ref{fig:essay_fig_47_1}, indicates the bias of this estimator $\hat{\varphi}_l$ exists when the actual value of $\varphi_i$ is fixed or dynamic. Therefore, it requires improvement of the analysis methods to achieve better estimation.

\subsection{Simulation study \RNum{3} - Dynamic parameters in different ranges}
In this section of study \RNum{3}, the outcomes of modelling and parameter estimation illustrate that the improvement of previous statistical methods impact estimation accuracy.

\begin{figure}
     \centering
     \begin{subfigure}[b]{0.49\textwidth}
         \centering
         \includegraphics[width=\textwidth]{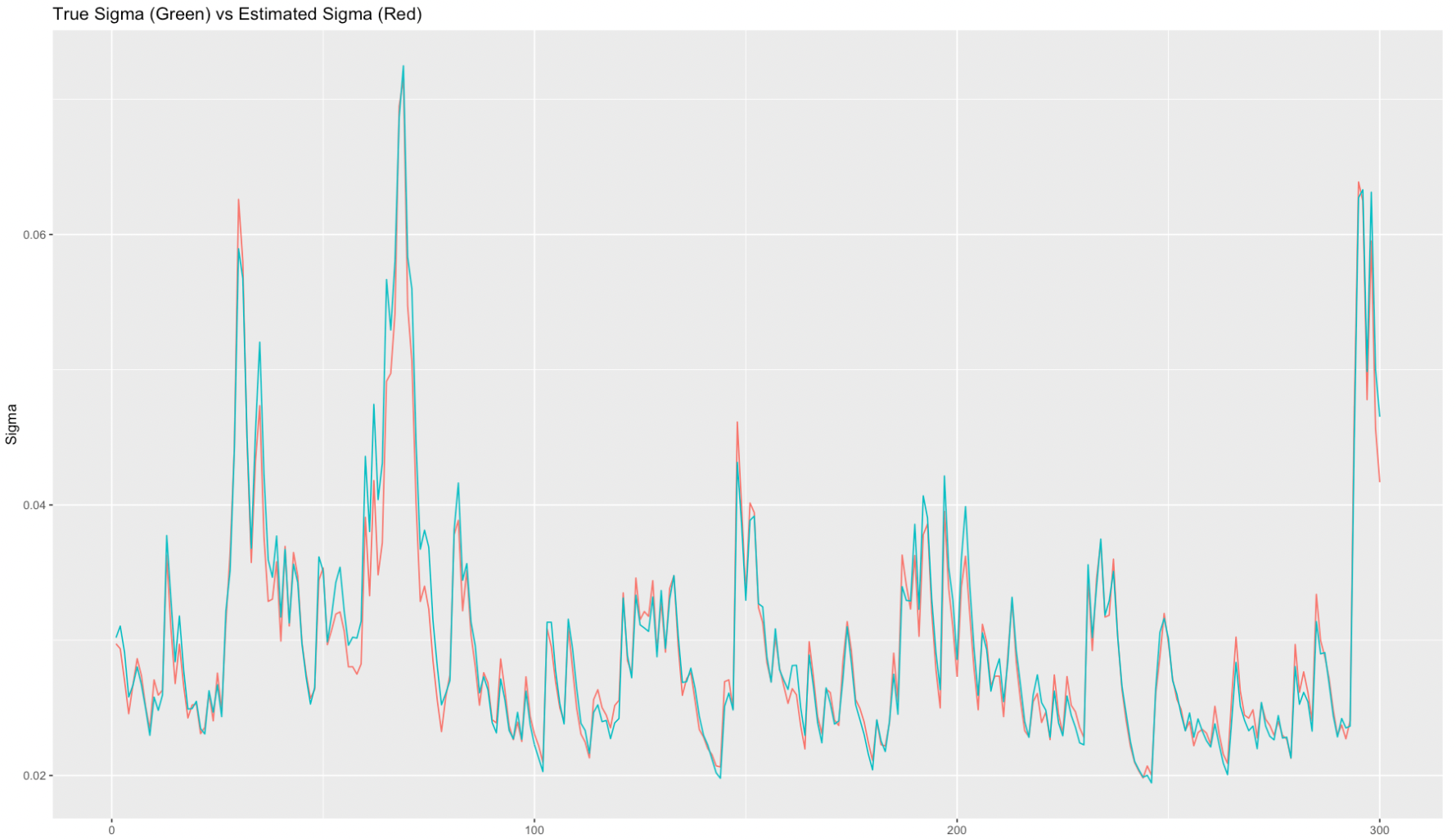}
         \caption{}
         \label{fig:essay_fig_48_1}
     \end{subfigure}
     \begin{subfigure}[b]{0.49\textwidth}
         \centering
         \includegraphics[width=\textwidth]{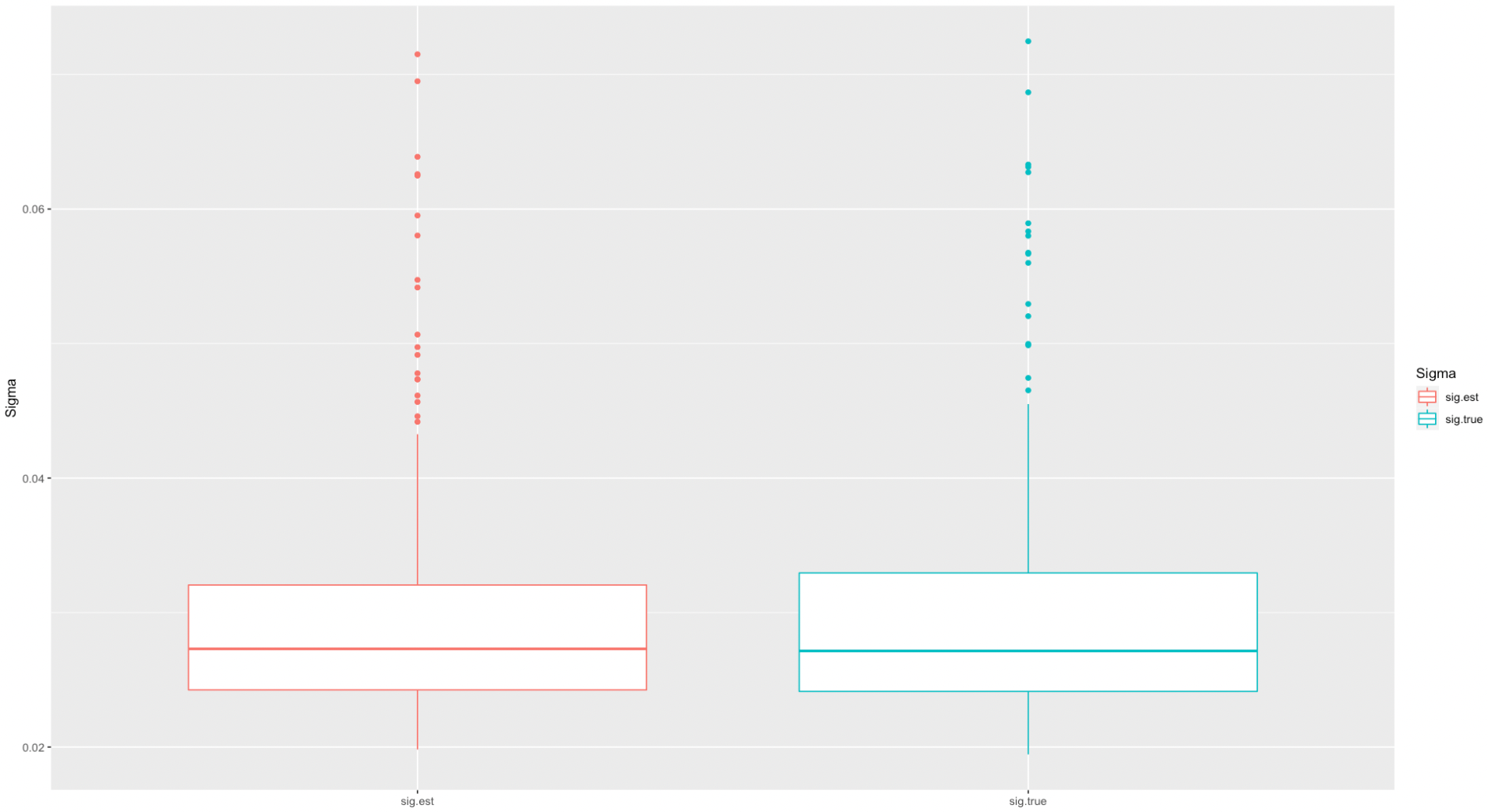}
         \caption{}
         \label{fig:essay_fig_48_2}
     \end{subfigure}
    \caption{Left plot shows true $\varphi_i$ and estimated $\hat{\varphi}_l$, right boxplot shows mean and variance of $\hat{\varphi}_l$ and $\varphi_i$ respectively}
    \label{fig:phi_mixplots}
\end{figure}

The first parameter to compare is $\sigma_{t \mid t-1}$ ($\hat{\sigma}_{t \mid t-1}$). The data set consists of 200 simulated time series with $\varphi_i$ close to 0.0 and 200 simulated time series with $\varphi_i$ close to 1.0. After fitting AR models on each series and computing the averaged residuals $\bar{\hat{\eta}}_t$ with weights $W_i$, the estimator $\hat{\sigma}_{t \mid t-1}$ is still close to $\sigma_{t \mid t-1}$. Although Figure \ref{fig:phi_mixplots} displays the bias of estimator $\hat{\sigma}_{t \mid t-1}$ on local minimum values and local maximum values on the left line plot, the boxplot on the right indicates the mean value of estimator $\hat{\sigma}_{t \mid t-1}$ is equal to mean of $\sigma_{t \mid t-1}$. Also, the variance of estimation is approximately equal to 0.00081572.

The estimation of standardized residuals $\varepsilon_t$ is good as expected since the Q-Q plot of estimator $\hat{\varepsilon}_t$ follows a normal distribution without many differences from $\hat{\varepsilon}_t$ generated from previous methods. In other words, the estimator $\hat{\varepsilon}_t$ is not impacted by averaging residuals \{$\hat{\eta}_{it}$\} with weights $W_i$. This supports the assumption that the improvement of statistical methods reduces the bias of estimator $\hat{\varphi}_l$ without influencing estimator $\hat{\varepsilon}_t$.

The idea of the improved analysis methods is calculating the estimation of $\varphi_i$ by averaging $\hat{\varphi}_{1i}$ and $\hat{\varphi}_{2i}$. For this reason, the new estimator $\hat{\varphi}_l$ considers outputs from two estimators $\hat{\varphi}_{1i}$ and $\hat{\varphi}_{2i}$ generated from two different steps during modelling.
\begin{figure}
     \centering
     \begin{subfigure}[b]{0.49\textwidth}
         \centering
         \includegraphics[width=\textwidth]{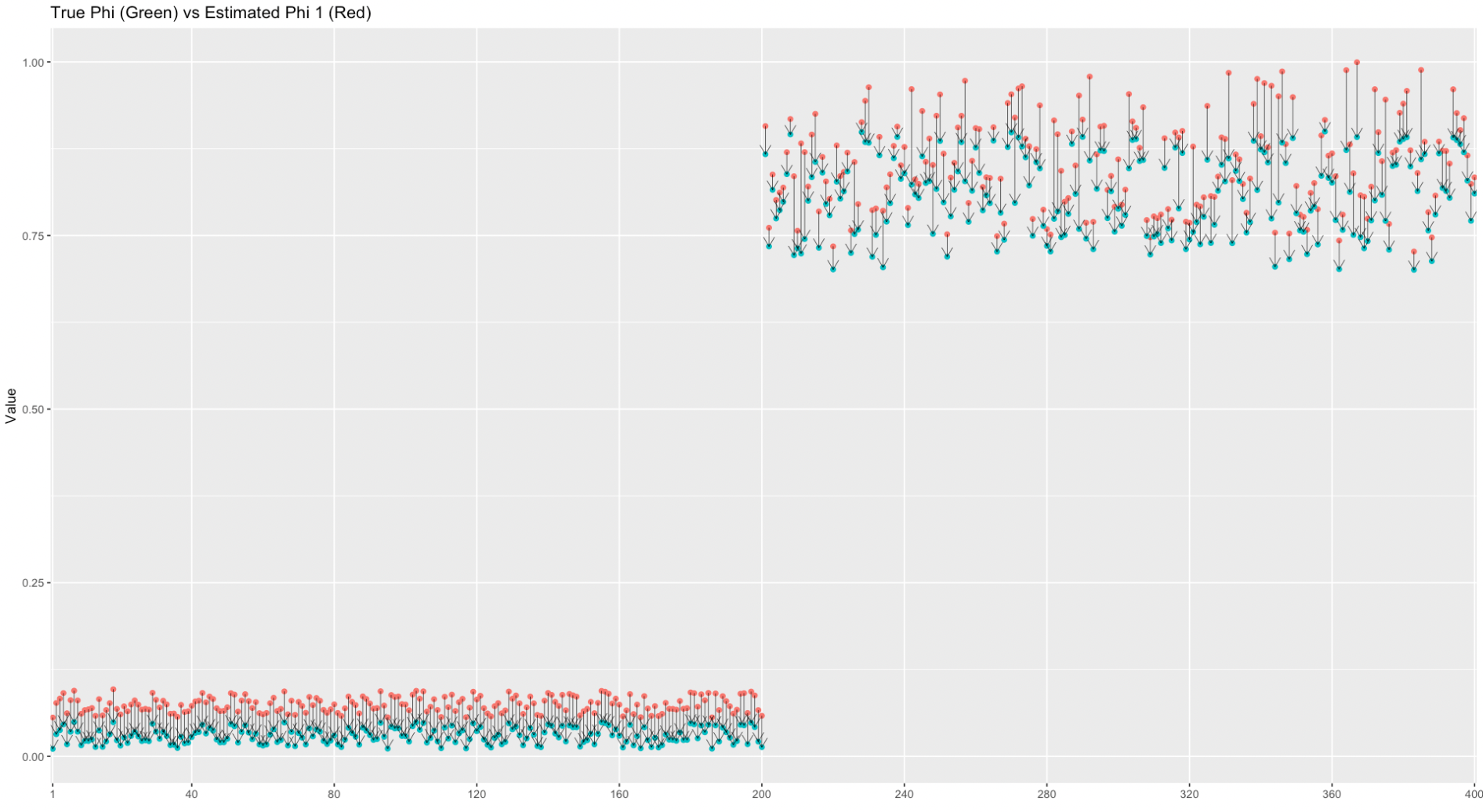}
         \caption{}
         \label{fig:essay_fig_49_1}
     \end{subfigure}
     \begin{subfigure}[b]{0.49\textwidth}
         \centering
         \includegraphics[width=\textwidth]{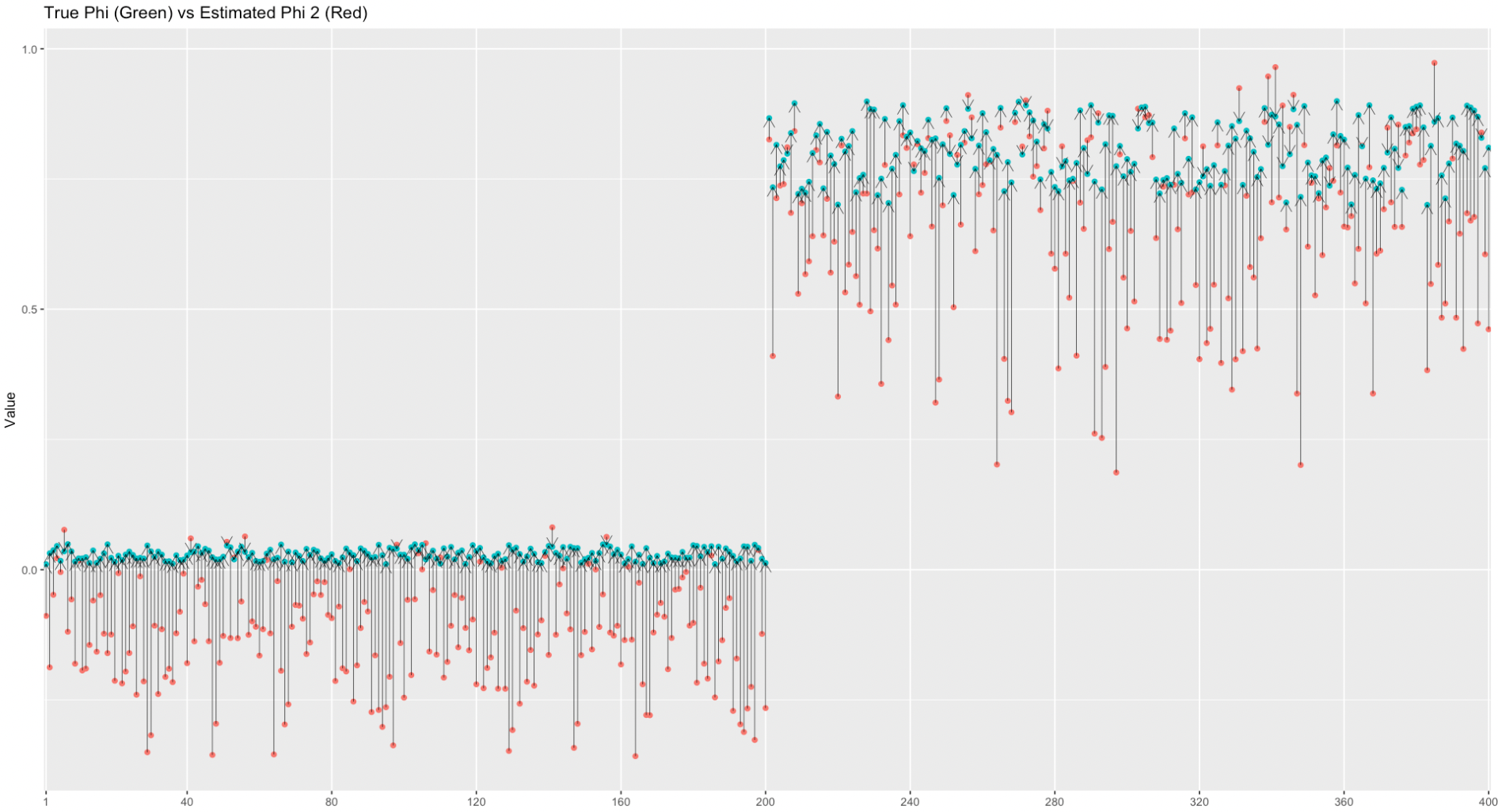}
         \caption{}
         \label{fig:essay_fig_49_2}
     \end{subfigure}
     \begin{subfigure}[b]{0.49\textwidth}
         \centering
         \includegraphics[width=\textwidth]{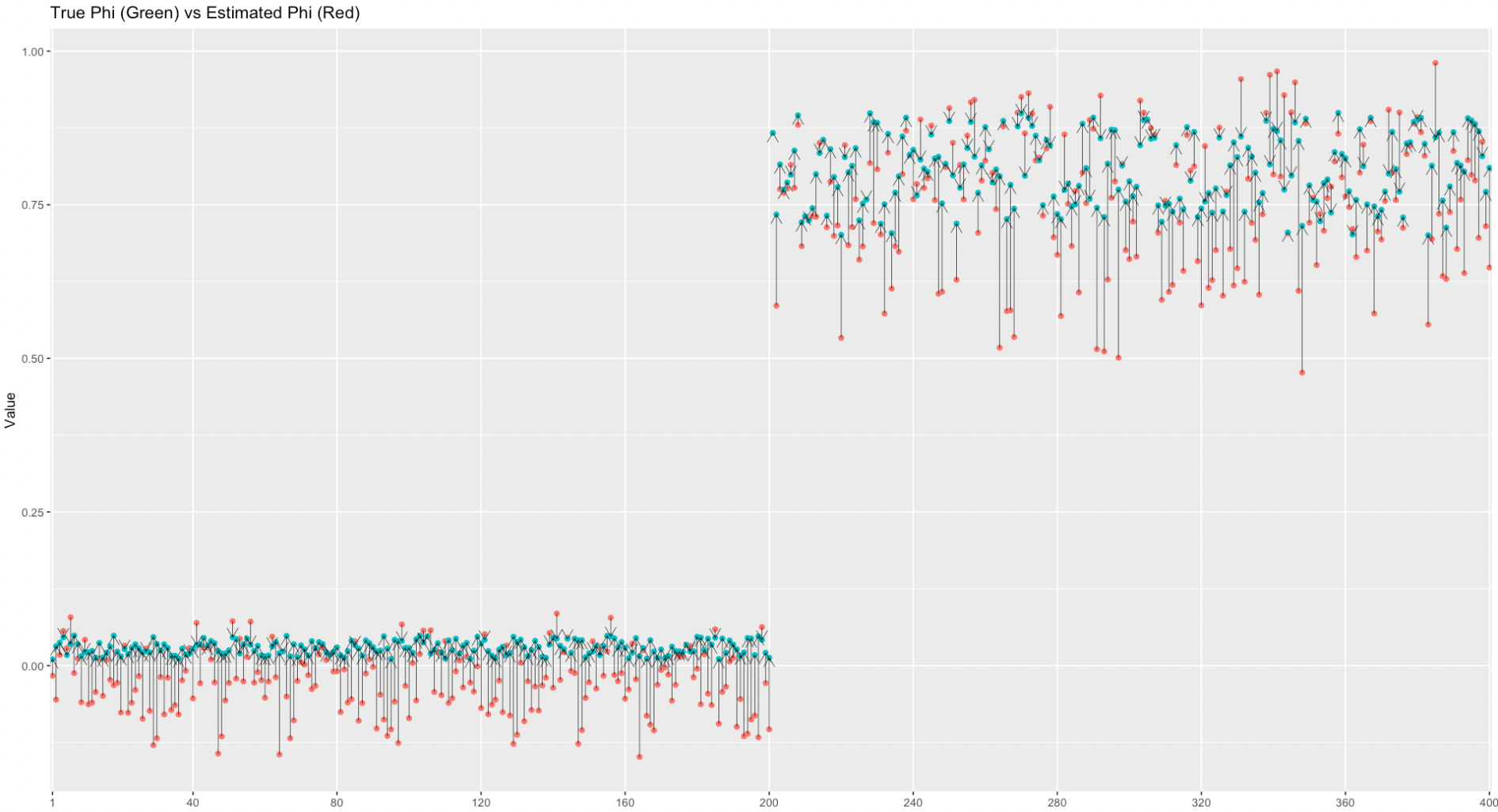}
         \caption{}
         \label{fig:essay_fig_49_3}
     \end{subfigure}
     \begin{subfigure}[b]{0.49\textwidth}
         \centering
         \includegraphics[width=\textwidth]{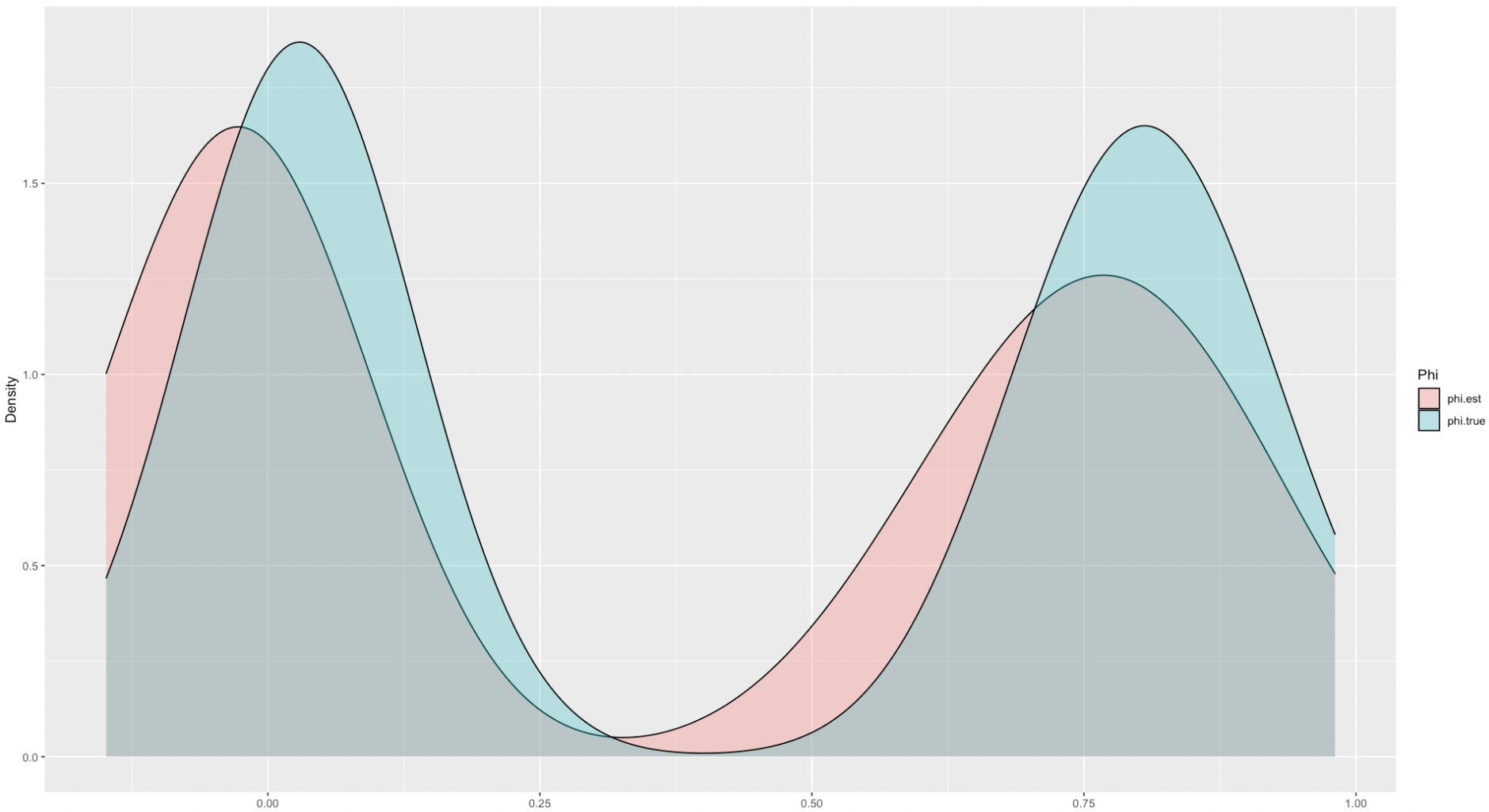}
         \caption{}
         \label{fig:essay_fig_49_4}
     \end{subfigure}     
    \caption{The top left plot shows bias of the first estimator $\hat{\varphi}_{1i}$, the top right plot shows bias of the second estimator $\hat{\varphi}_{2i}$. The bottom left plot indicates more accurate estimator $\hat{\varphi}_{l}$ after averaging $\hat{\varphi}_{1i}$ and $\hat{\varphi}_{2i}$ The bottom right plot shows density of estimator $\hat{\varphi}_{l}$ in two ranges ($0.01,0.05$) and ($0.7,0.9$)}
    \label{fig:essay_fig_49}
\end{figure}

The first estimator $\hat{\varphi}_{1i}$ is generated with bias after fitting AR model on each series. The top left plot in Figure \ref{fig:essay_fig_49} reveals that most estimation values are smaller than the actual values. Furthermore, the top right plot shows bias when most estimation values are larger than the actual values. Hence, the second estimator $\hat{\varphi}_{2i}$ also generates the bias. Whereas the way of averaging $\hat{\varphi}_{1i}$ and $\hat{\varphi}_{2i}$ reduce the bias of estimator $\hat{\varphi}_l$ remarkably, illustrated by the bottom left plot in Figure \ref{fig:essay_fig_49}. It is expected to achieve minimum bias of estimator $\hat{\varphi}_l$ without significantly influencing the accuracy of other estimators $\hat{\sigma}_{t \mid t-1}$ and $\hat{\varepsilon}_t$.
\begin{table}
    \centering
    \begin{tabular}{lllll}
        \toprule
        K & $\varphi=0.05$ & $0.01<\varphi<0.05$ & $0.7<\varphi<0.9$ & Results\\
        \midrule
        20 & 0.01736553 & - & - & 0.01736553 \\
        100 & 0.06222519 & - & - & 0.06222519 \\
        400 & 0.06175203 & - & - & 0.06175203 \\
        100 & - & - & 0.03460226 & 0.03460226 \\
        400 & - & - & 0.0364705 & 0.0364705 \\
        \midrule
        200 & - & \textbf{0.005858768} & - & \multirow{2}{*}{\textbf{0.01736553}} \\
        200 & - & - & \textbf{0.009149003} &  \\
        \bottomrule
    \end{tabular}
    \caption{Summary of estimation results for parameter $\varphi_i$ in different ranges with different sizes of data sets}
    \label{tab:table_43}
\end{table}

Table \ref{tab:table_43} summarises the estimation results of $\varphi_i$ that are achieved when $\varphi_i$ is fixed and dynamic in different ranges, ($0.01,0.05$) and ($0.7,0.9$), with varying sizes of simulated data sets. It demonstrates that the improved method with proposed weights parameter $W_i$ avoids the apparent bias of estimator $\hat{\varphi}_l$ as expected. It is necessary to calculate the estimator $\hat{\varphi}_l$ twice to give the final estimation of $\varphi_i$.

\subsection{Practice study – Real-world fMRI data}
The following steps are performed to model the real-world fMRI data: (1) Identify AR orders $\bar{u}$ of each fMRI series and fit estimated AR ($\bar{u}$) model to get estimator $\hat{\varphi}_{1i}$ and residuals \{$\hat{\eta}_{it}$\}. (2) Use estimator $\hat{\varphi}_{1i}$ to calculate weights $W_i$, then calculate averaged residuals $\bar{\hat{\eta}}_t$ with the equation: $\bar{\hat{\eta}}_t = \sum W_i \hat{\eta}_{it} $. (3) Identify GARCH orders ($\hat{p},\hat{q}$) on $\bar{\hat{\eta}}_t$ and fit a GARCH($\hat{p},\hat{q}$) model. (4) Without averaging residuals, fit each residuals $\eta_{it}$ with GARCH($\hat{p},\hat{q}$) models to get estimator $\hat{\varphi}_{i\_old}$ (this is the old analysis method used for estimating $\varphi_i$). (5) Remove $\bar{\eta}_t$ from each fMRI series and fit returns with AR($\bar{u}$) model to get estimator $\hat{\varphi}_{2i}$. (6) Calculate estimator $\hat{\varphi}_i$ by averaging $\hat{\varphi}_{1i}$ and $\hat{\varphi}_{2i}$. Statically assess GARCH model fitting and compare parameter estimators $ \hat{\varphi}_i$ ($= (\hat{\varphi}_{1i} + \hat{\varphi}_{2i}) / 2 $) to $\hat{\varphi}_{i\_old}$.

\paragraph{Subject CC10056:} The findings of our study on CC10056 are presented in this section.

\begin{figure}
     \centering
     \begin{subfigure}[b]{0.49\textwidth}
         \centering
         \includegraphics[width=\textwidth]{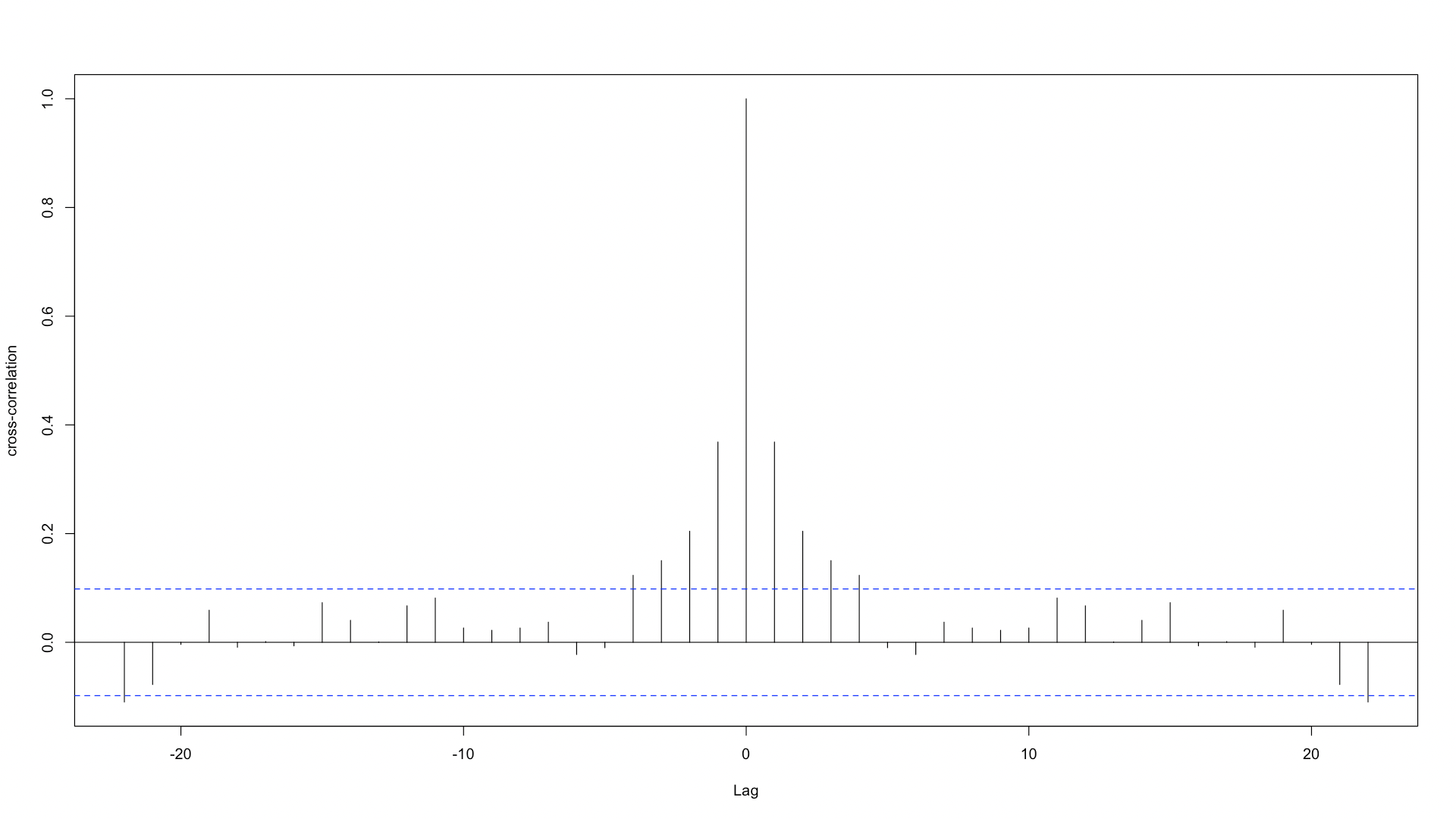}
         \caption{}
         \label{fig:essay_fig_410_1}
     \end{subfigure}
     \begin{subfigure}[b]{0.49\textwidth}
         \centering
         \includegraphics[width=\textwidth]{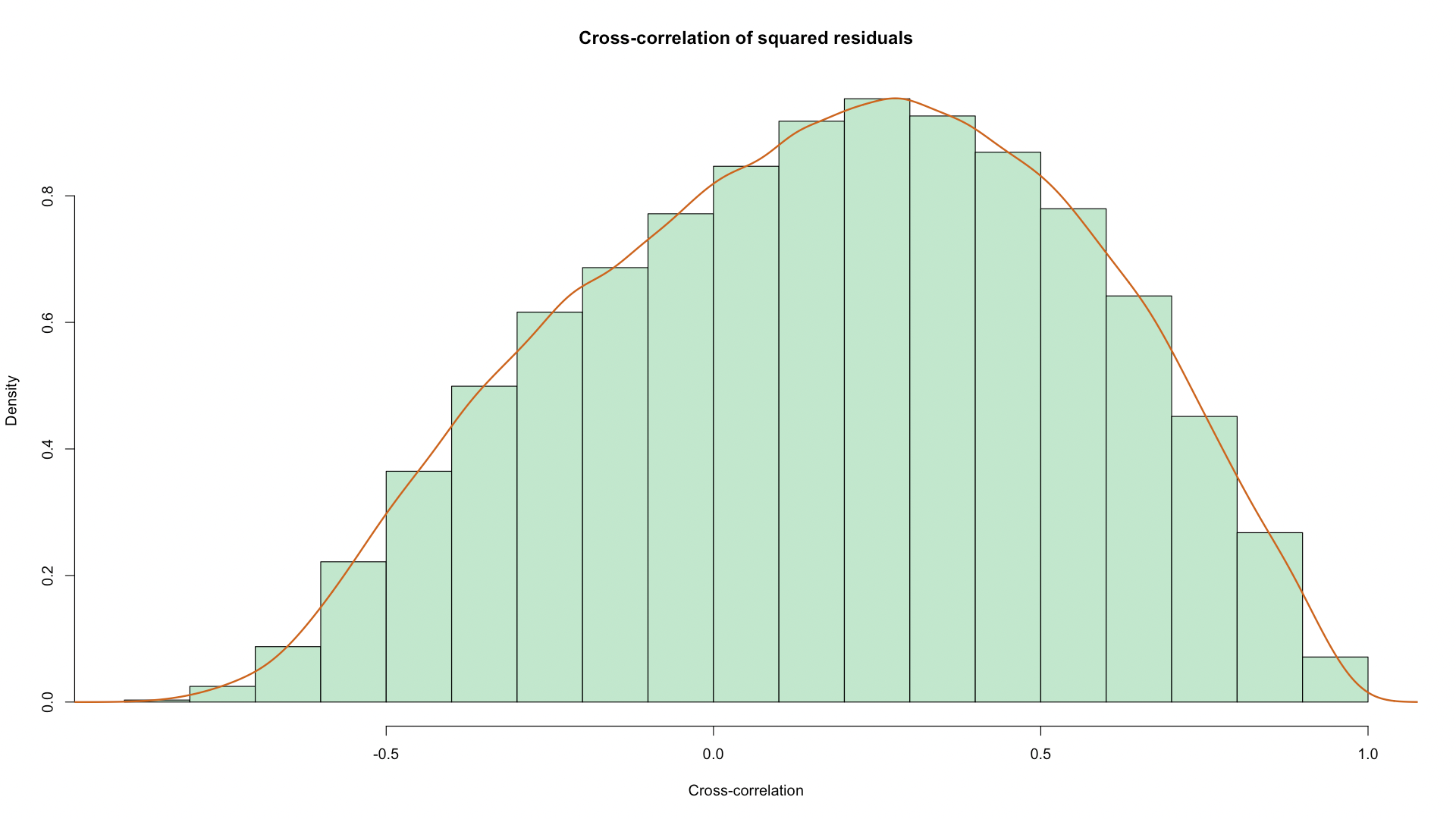}
         \caption{}
         \label{fig:essay_fig_410_2}
     \end{subfigure}
    \caption{The left plot shows cross-correlation of \{$\hat{\eta}_{it}$\}. The density of cross-correlation on the right plot}
    \label{fig:essay_fig_410}
\end{figure}

It is necessary to justify the GARCH model choice that the volatility is shared over the 400 ROIs. Accordingly, the $400 \times 400$ cross-correlation matrix between the squared residuals \{$\hat{\eta}_{it}$\} is calculated and plotted visually. Figure \ref{fig:essay_fig_410} shows that the distribution of cross-correlation is not centred on zero, instead it is centred on a positive value.

The plot of averaged residuals $\bar{\hat{\eta}}_t$ clearly reveals volatility clustering. Also, it fails McLeod-Li, so there is evidence of ARCH type behaviour in the model.
\begin{figure}
     \centering
     \begin{subfigure}[b]{0.49\textwidth}
         \centering
         \includegraphics[width=\textwidth]{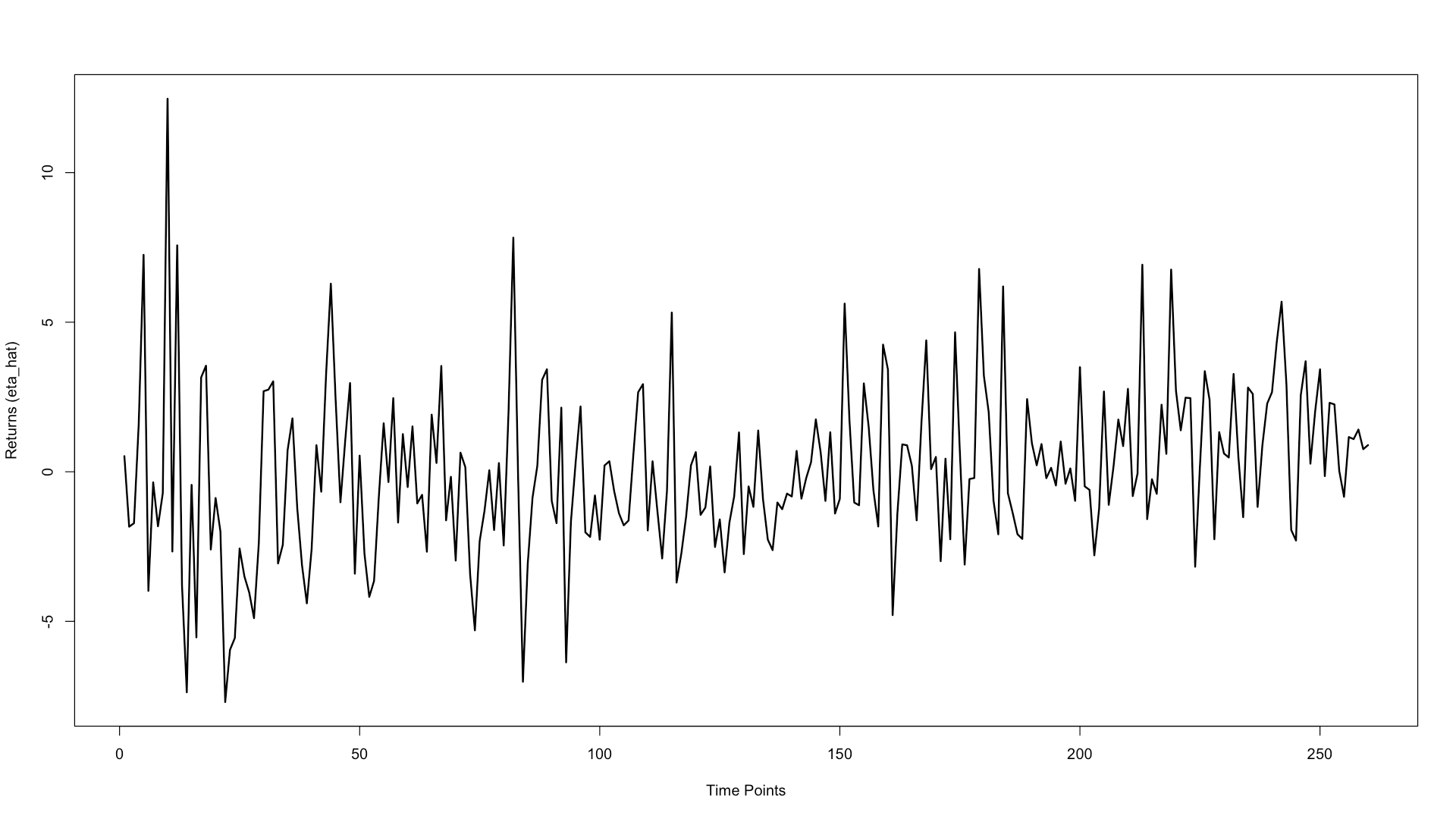}
         \caption{}
         \label{fig:essay_fig_411_1}
     \end{subfigure}
     \begin{subfigure}[b]{0.49\textwidth}
         \centering
         \includegraphics[width=\textwidth]{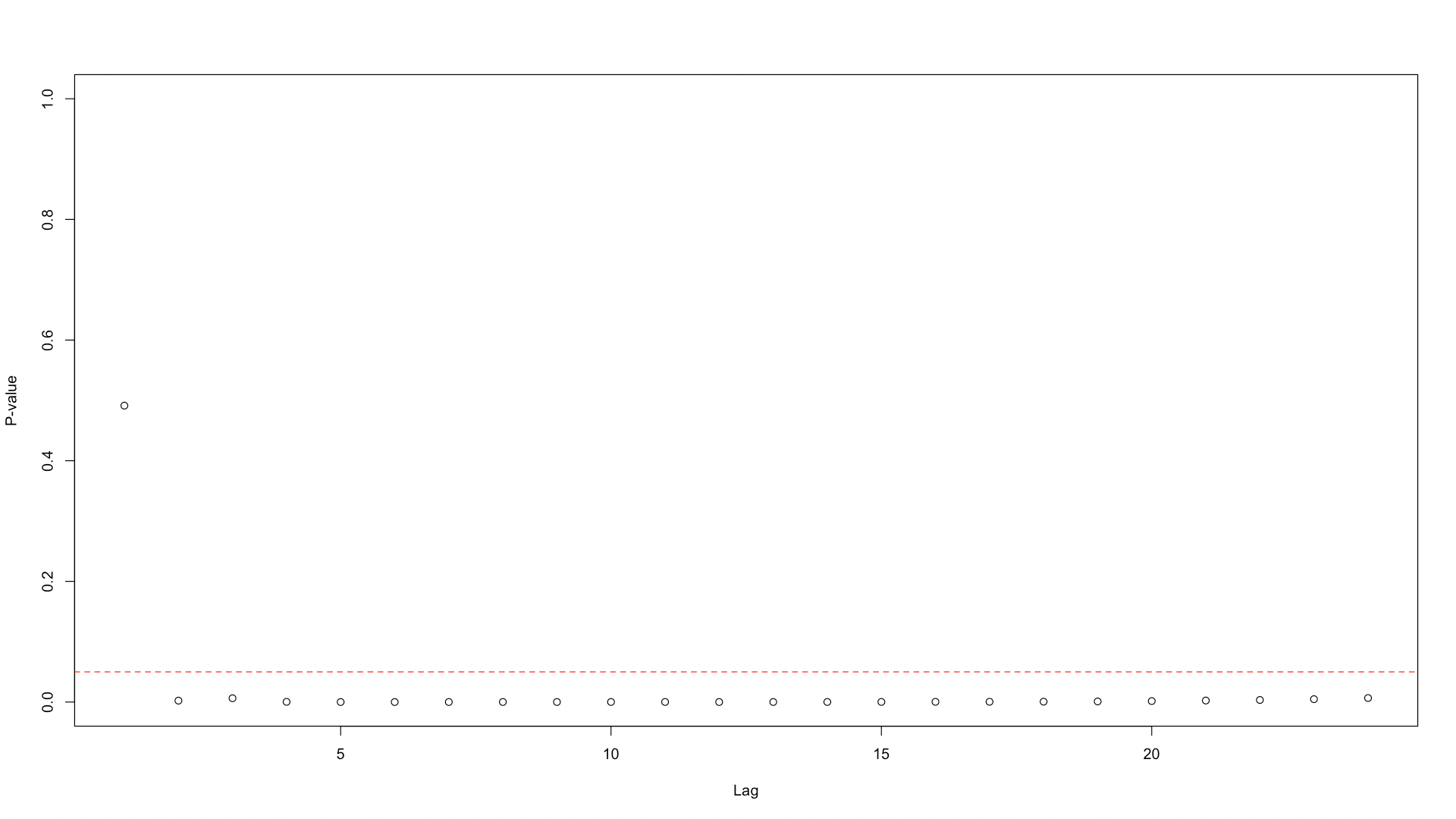}
         \caption{}
         \label{fig:essay_fig_411_2}
     \end{subfigure}
    \caption{The left plot shows volatility clustering. The McLeod-Li test results on the right}
    \label{fig:essay_fig_411}
\end{figure}

The averaged residuals $\bar{\hat{\eta}}_t$ can be fitted with a GARCH($1,1$) model. The coefficients $\hat{\alpha}_1$ is significant, although $\hat{\beta}_1$ is slightly larger than 0.05 if the confidence interval is considered 0.05.
\begin{table}
    \centering
    \begin{tabular}{lllll}
        \toprule
        & Estimate & Std. Error & t value & Pr ($ > \mid t \mid $)\\
        \midrule
        omega & $0.52535$ & $0.34033$ & $1.544$ & $0.1227$ \\
        alpha1 & $0.06399$ & $0.03611$ & $1.772$ & $0.0764$ \\
        beta1 & $0.86132$ & $0.06874$ & $12.529$ & $<2e-16$ \\
        \bottomrule
    \end{tabular}
    \caption{Summary of estimated coefficients for GARCH model on CC10056 shared residuals}
    \label{tab:table_44}
\end{table}

Another model, GARCH($2,2$), is also fitted on $\bar{\hat{\eta}}_t$ to compare fitting results to GARCH($1,1$). The AIC indicates that GARCH($1,1$) model is better than GARCH($2,2$) in terms of fitting $\bar{\hat{\eta}}_t$.
\begin{table}
    \centering
    \begin{tabular}{lll}
        \toprule
        & GARCH($1,1$) & GARCH($2,2$) \\
        \midrule
        AIC & 4.834117 & 4.844648 \\
        \bottomrule
    \end{tabular}
    \caption{AIC for GARCH($1,1$) and GARCH($2,2$)}
    \label{tab:table_45}
\end{table}

Therefore, the equation of the GARCH model is written as below:
\begin{equation}
\begin{split}
    \bar{\eta}_t &= \sigma_{t \mid t-1} \varepsilon_t \\
    \sigma^2_{t \mid t-1} &= 0.52535 + 0.06399\eta^2_{t-1} + 0.86132\sigma^2_{t-1 \mid t-2} \\
    \varepsilon_t &\sim N(0,1) \\
\end{split}
\end{equation}

The final line shows the Li-Mak test (Li and Mak, 1994) results, indicating that we successfully modelled out the ARCH behaviour in the series. Equivalently, the Li-Mak test inspects the presence of autocorrelation in their squares, showing a sign that the GARCH model captures all autoregressive conditional heteroskedastic patterns there are.

\begin{figure}
     \centering
     \begin{subfigure}[b]{0.49\textwidth}
         \centering
         \includegraphics[width=\textwidth]{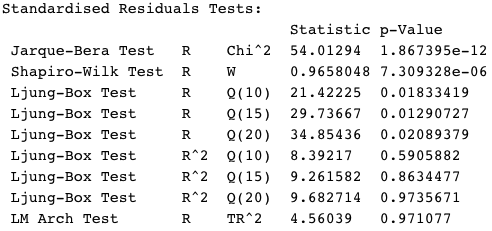}
         \caption{}
         \label{fig:essay_fig_412_1}
     \end{subfigure}
     \begin{subfigure}[b]{0.49\textwidth}
         \centering
         \includegraphics[width=\textwidth]{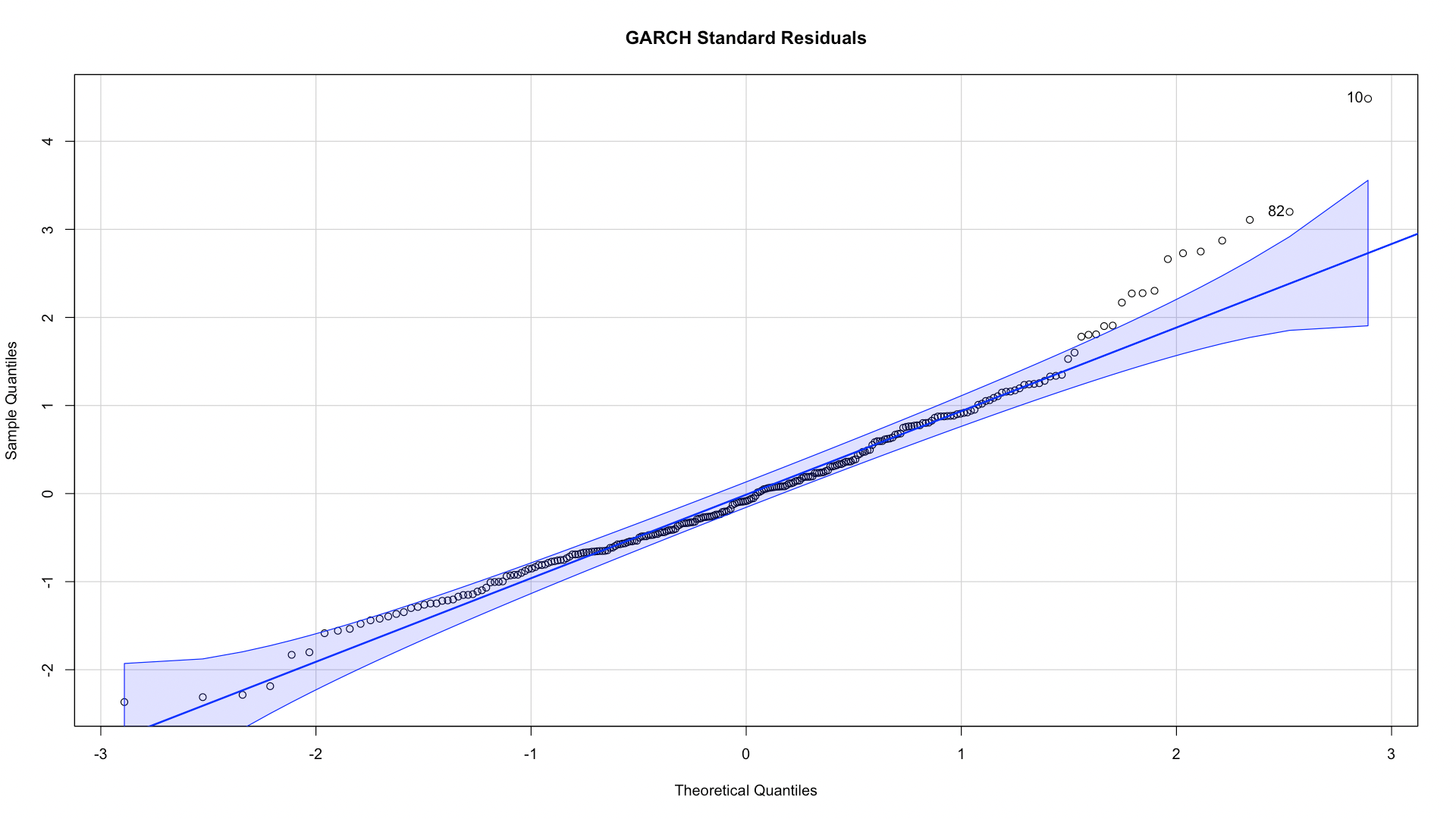}
         \caption{}
         \label{fig:essay_fig_412_2}
     \end{subfigure}
    \caption{The left plot shows tests for ARCH/GARCH behaviour in standardized residuals The right Q-Q plot of standardized residuals reveals a normal distribution}
    \label{fig:essay_fig_412}
\end{figure}

The standard residuals $\hat{\varepsilon}_t$ is examined by a Q-Q plot with the confidence interval of $0.05$. It illustrates standardized residuals follow a normal distribution, see Figure \ref{fig:essay_fig_412}.

The estimator $\hat{\varphi}_{1i}$, $\hat{\varphi}_{2i}$ and $\hat{\varphi}_{i\_old}$ are examined by scatter plots. Since some fMRI series are identified as AR models with orders larger than 2, in this case, only the first two coefficients $\hat{\varphi}_1$ and $\hat{\varphi}_2$ are discussed. To compare the estimator $\hat{\varphi}_i$ (generated from the improved analysis method) to the estimator $\hat{\varphi}_{i\_old}$ (generated from old analysis method), a possible method is comparing their standard error returned from modelling. Remember that estimator $\hat{\varphi}_i$ is equal to $(\hat{\varphi}_{1i}$ + $\hat{\varphi}_{2i})/2$, thus the standard error of $\hat{\varphi}_i$ is given by the following equation:
\begin{equation}
\begin{split}
    \operatorname{SE}(\hat{\varphi}_i) &= \sqrt{\operatorname{Var}(\hat{\varphi}_i)} \\
    &= \sqrt{\frac{1}{4}\operatorname{Var}(\hat{\varphi}_{1i} + \hat{\varphi}_{2i})} \\
    &= \frac{1}{2}\sqrt{\operatorname{Var}(\hat{\varphi}_{1i}) + \operatorname{Var}(\hat{\varphi}_{2i})} \\
\end{split}
\end{equation}

\begin{figure}
     \centering
     \begin{subfigure}[b]{0.49\textwidth}
         \centering
         \includegraphics[width=\textwidth]{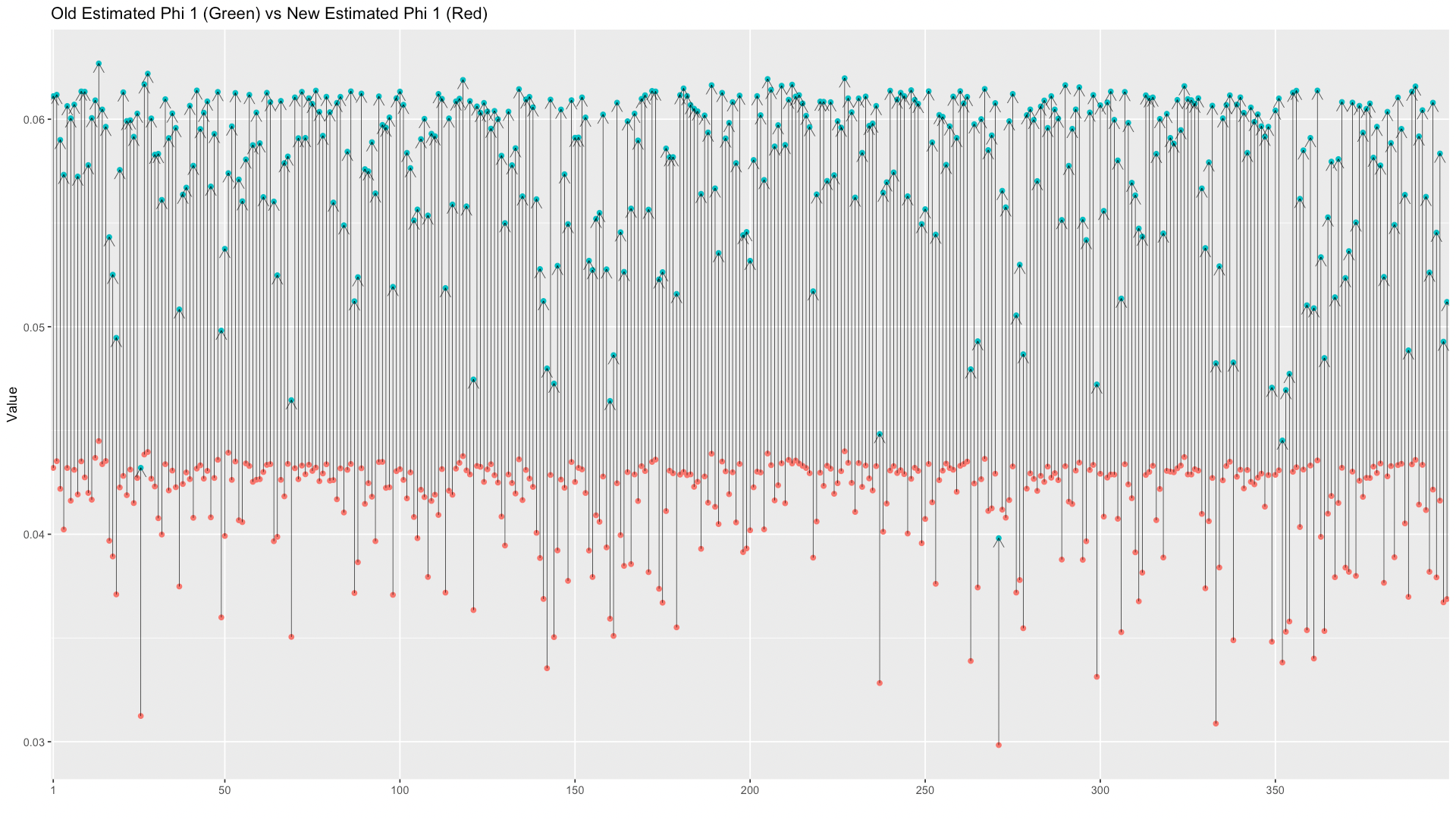}
         \caption{}
         \label{fig:essay_fig_413_1}
     \end{subfigure}
     \begin{subfigure}[b]{0.49\textwidth}
         \centering
         \includegraphics[width=\textwidth]{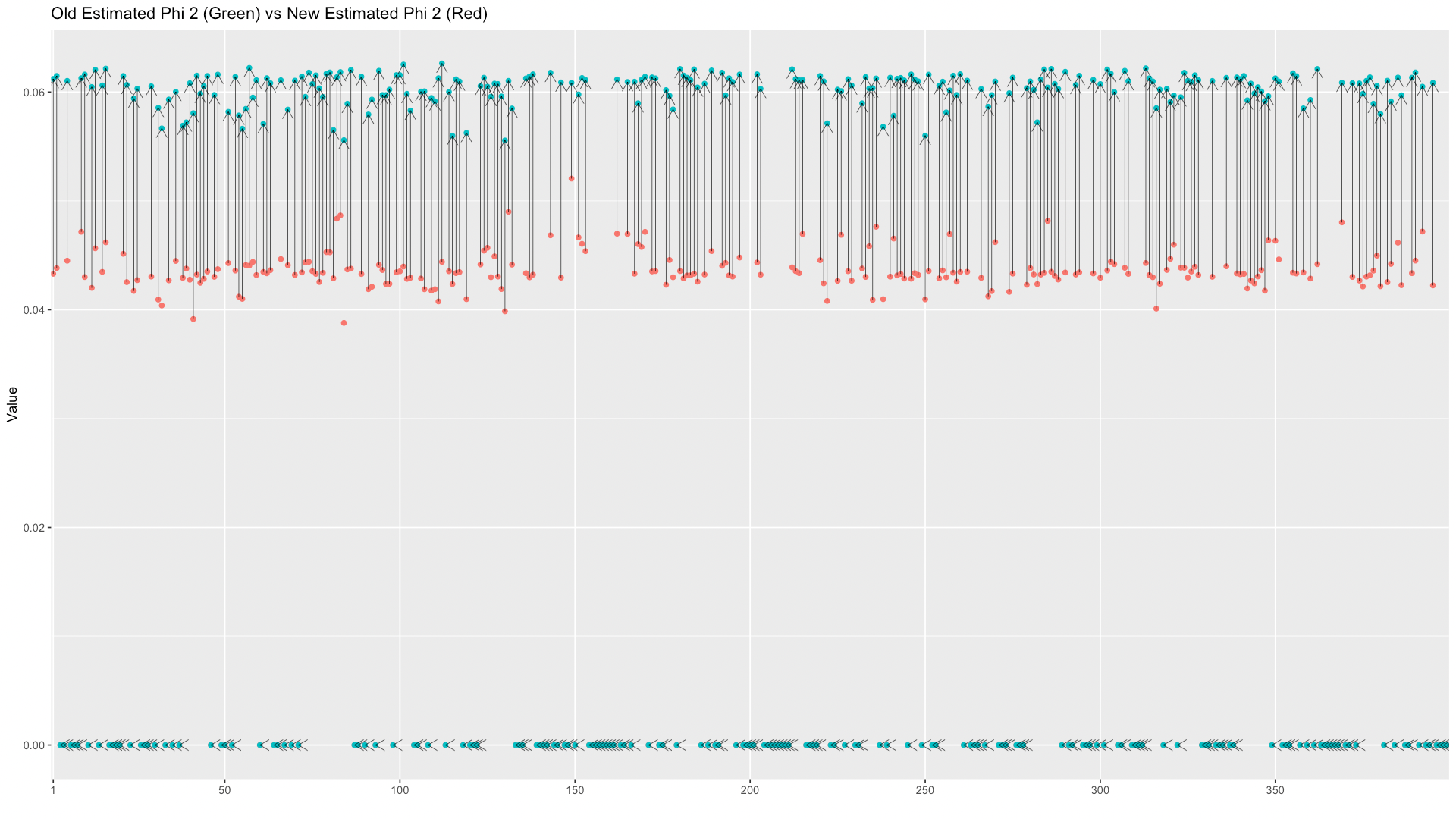}
         \caption{}
         \label{fig:essay_fig_413_2}
     \end{subfigure}
    \caption{The left plot standard error of estimation of the first coefficient $\varphi_1$. The right plot standard error of estimation of the second coefficient $\varphi_2$}
    \label{fig:essay_fig_413}
\end{figure}

The variance of $\hat{\varphi}_{1i}$ and $\hat{\varphi}_{2i}$ can be calculated by squaring off the standard error of $\hat{\varphi}_{1i}$ and $\hat{\varphi}_{2i}$ returned by AR model. Using visualization tools is the straight method to compare two estimators for the first coefficient $\varphi_1$ and the second coefficient $\varphi_2$. The scatter plots, Figure \ref{fig:essay_fig_413} reveal that the standard error of the estimator $\hat{\varphi}_i$ is smaller than the standard error of old estimator $\hat{\varphi}_{i\_old}$ if the coefficients $\hat{\varphi}_i$ and $\hat{\varphi}_{i\_old}$ are neither equal to zero.

\begin{figure}
     \centering
     \begin{subfigure}[b]{0.3\textwidth}
         \centering
         \includegraphics[width=\textwidth]{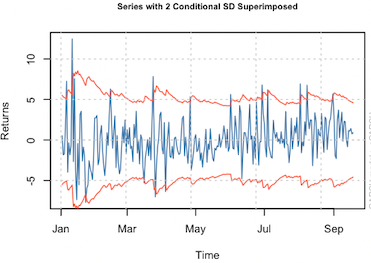}
         \caption{}
         \label{fig:essay_fig_414_1}
     \end{subfigure}
     \begin{subfigure}[b]{0.3\textwidth}
         \centering
         \includegraphics[width=\textwidth]{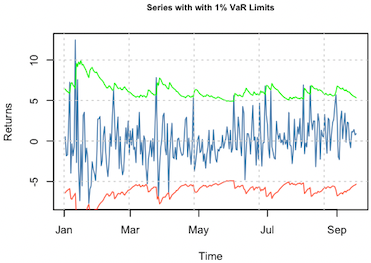}
         \caption{}
         \label{fig:essay_fig_414_2}
     \end{subfigure}
     \begin{subfigure}[b]{0.3\textwidth}
         \centering
         \includegraphics[width=\textwidth]{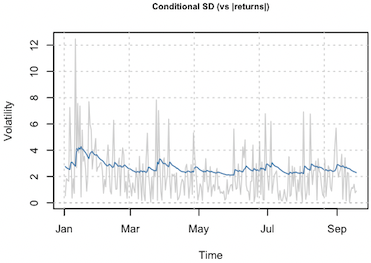}
         \caption{}
         \label{fig:essay_fig_414_3}
     \end{subfigure}     
     \begin{subfigure}[b]{0.3\textwidth}
         \centering
         \includegraphics[width=\textwidth]{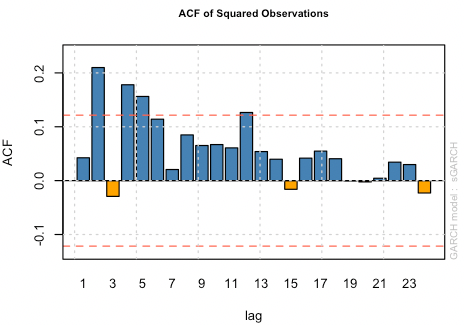}
         \caption{}
         \label{fig:essay_fig_414_4}
     \end{subfigure}
     \begin{subfigure}[b]{0.3\textwidth}
         \centering
         \includegraphics[width=\textwidth]{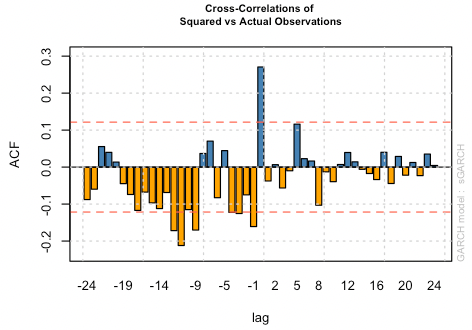}
         \caption{}
         \label{fig:essay_fig_414_5}
     \end{subfigure} 
     \begin{subfigure}[b]{0.3\textwidth}
         \centering
         \includegraphics[width=\textwidth]{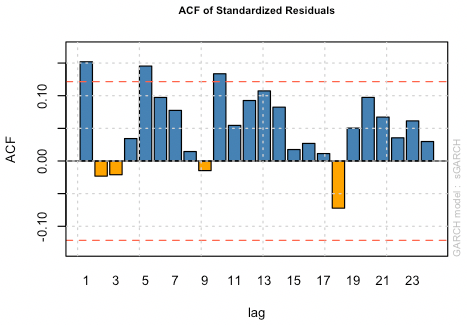}
         \caption{}
         \label{fig:essay_fig_414_6}
     \end{subfigure}     
    \caption{(a) and (b) show the residuals $\bar{\hat{\eta}}_t$ against conditional variance. (c) plots conditional standard deviation against residuals $\bar{\hat{\eta}}_t$. (d) ACF of squared $\bar{\hat{\eta}}_t$. (e) Cross-correlation of $\bar{\hat{\eta}}_t$ and squared $\bar{\hat{\eta}}_t$. (f) ACF of standardized residuals $\hat{\varepsilon}_t$}
    \label{fig:essay_fig_414}
\end{figure}

When averaged residuals $\bar{\hat{\eta}}_t$ and conditional variance $\hat{\sigma}_{t \mid t-1}$ are displayed together, Figure \ref{fig:essay_fig_414_1} and \ref{fig:essay_fig_414_2}, the GARCH($1,1$) model tries to capture volatility. It mainly evidences that the model tracks some significant volatility clusters well. Moreover, Figure \ref{fig:essay_fig_414_4}, the ACF plot of squared $\bar{\hat{\eta}}_t$ shows high cross-correlation, and the ACF of standardized residuals $\hat{\varepsilon}_t$, Figure \ref{fig:essay_fig_414_6}, does not display many autocorrelations at low lags.

\paragraph{Subject CC10045}
The findings of our study on CC10045 are presented in this section.

\begin{figure}
     \centering
     \begin{subfigure}[b]{0.49\textwidth}
         \centering
         \includegraphics[width=\textwidth]{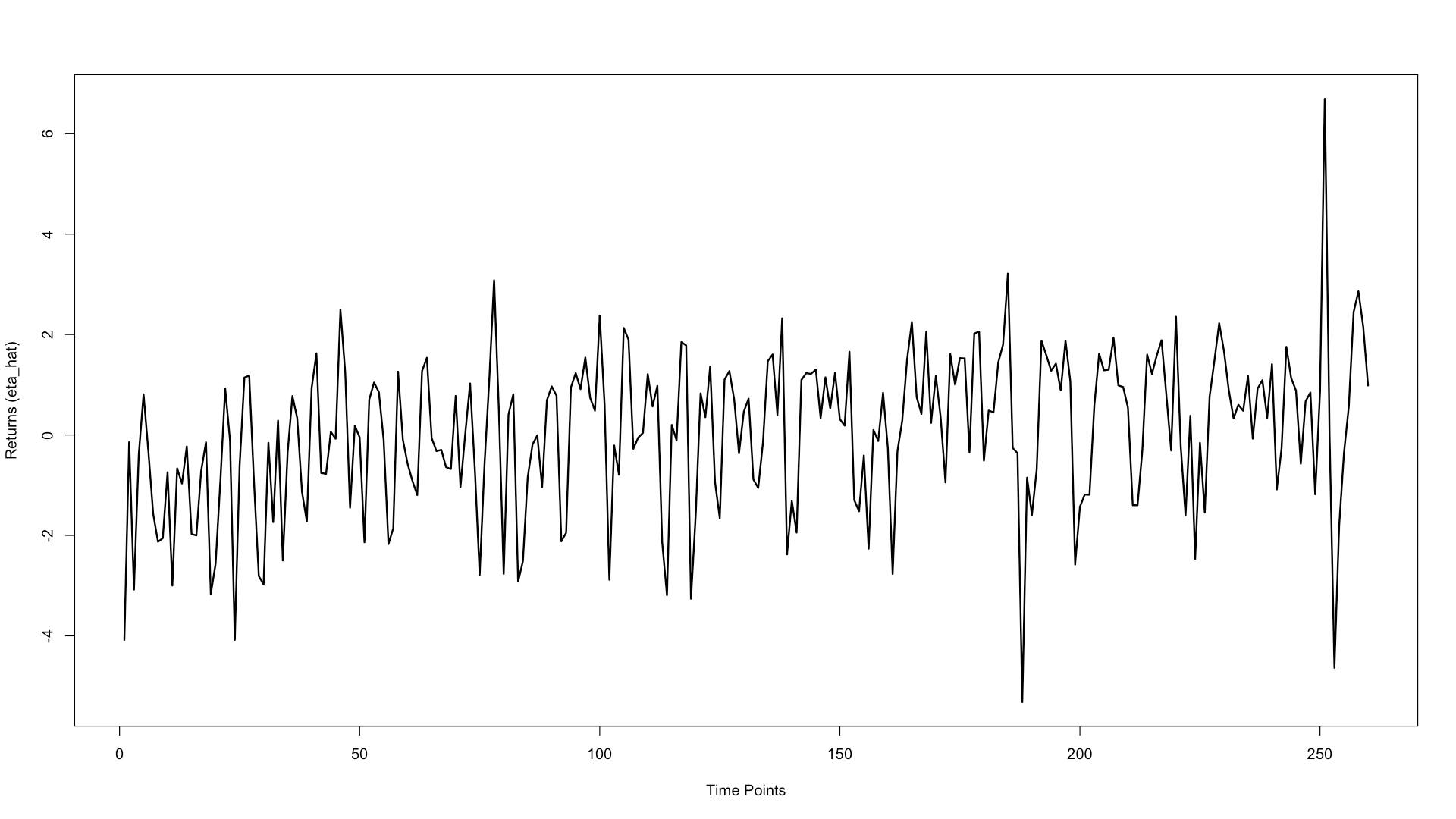}
         \caption{}
         \label{fig:essay_fig_415_1}
     \end{subfigure}
     \begin{subfigure}[b]{0.49\textwidth}
         \centering
         \includegraphics[width=\textwidth]{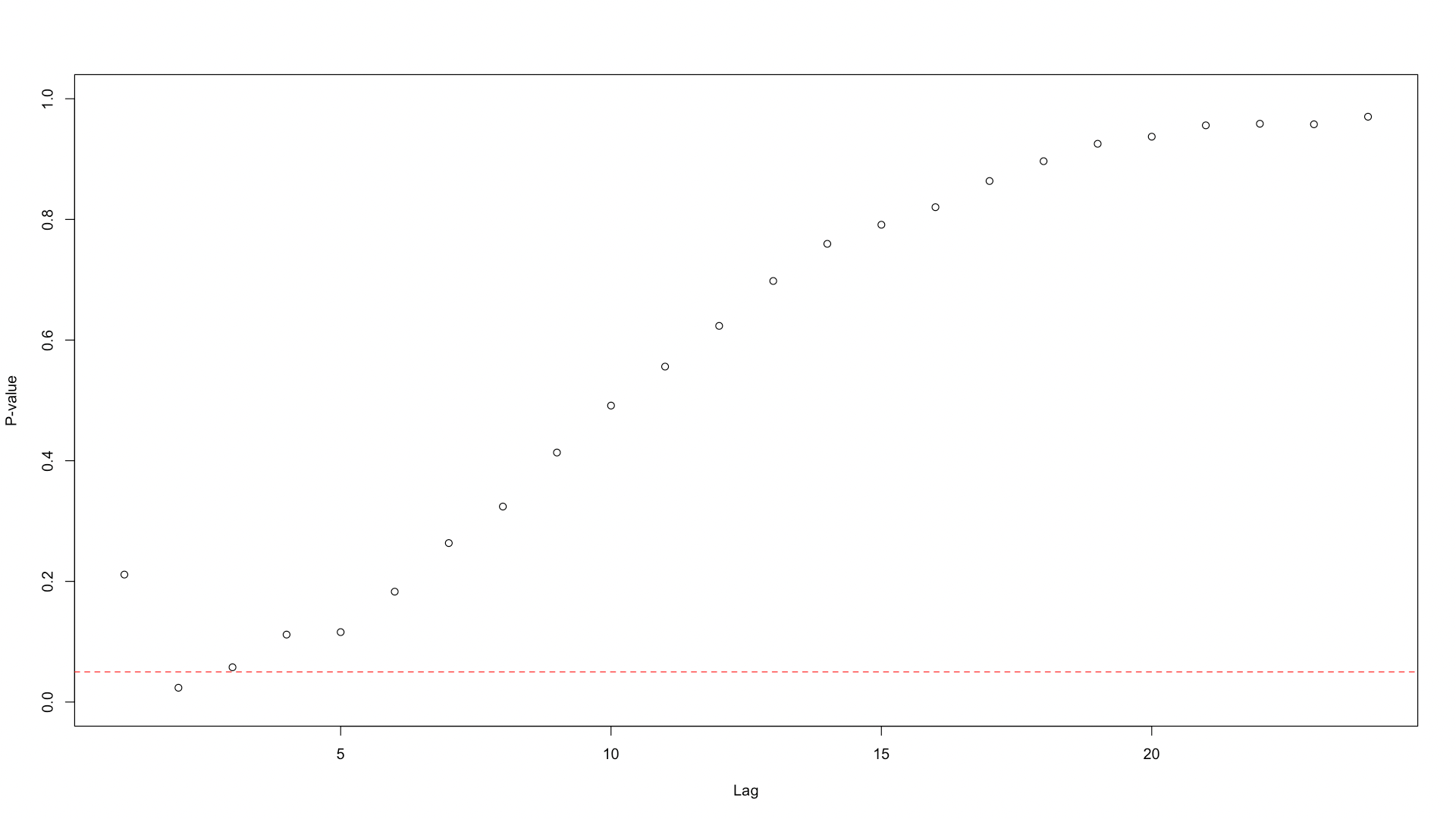}
         \caption{}
         \label{fig:essay_fig_415_2}
     \end{subfigure}
    \caption{The left plot shows volatility clustering. The McLeod-Li test results on the right}
    \label{fig:essay_fig_415}
\end{figure}

The plot of averaged residuals $\bar{\hat{\eta}}_t$ does not show much high volatility. Instead, it shows volatility clustering at the last few time points. Also, it passed McLeod-Li test, so the null hypothesis cannot be rejected. Thus there is no evidence of ARCH type behaviour in the model, see Figure \ref{fig:essay_fig_415}.

It is still necessary to fit the averaged residuals $\bar{\hat{\eta}}_t$ with a GARCH model to assess its parameters statistically. The orders of GARCH are identified as ($2,1$). A GARCH($2,1$) model is fitted. The coefficients $\hat{\beta}_1$ is significant if the confidence interval is considered $0.05$. But $\hat{\alpha}_1$ and $\hat{\alpha}_2$ have NaN p-values, which means their fitted probabilities are 0 or 1. The p-values suggest that only $\hat{\beta}_1$ is the significant coefficient required. Nevertheless, building any GARCH or ARCH model with coefficients $\hat{\beta}_1$ only is impossible, see Table \ref{tab:table_46}.
\begin{table}
    \centering
    \begin{tabular}{lllll}
        \toprule
        & Estimate & Std. Error & t value & Pr ($ > \mid t \mid $)\\
        \midrule
        omega & $6.090e-01$ & $4.849e-01$ & $1.256$ & $0.20910$ \\
        alpha1 & $1.000e-08$ & $NaN$ & $NaN$ & $NaN$ \\
        alpha2 & $5.550e-02$ & $NaN$ & $NaN$ & $NaN$ \\
        beta1 & $6.855e-01$ & $2.094e-01$ & $3.273$ & $0.00106$ \\
        \bottomrule
    \end{tabular}
    \caption{Summary of estimated coefficients for GARCH($2,1$) model on CC10045 shared residuals}
    \label{tab:table_46}
\end{table}
Despite this, other models, GARCH($1,1$) and GARCH($2,2$), are still fitted on $\bar{\hat{\eta}}_t$ to compare fitting results to GARCH($2,1$). The AIC, in Table \ref{tab:table_47}, indicates that GARCH($2,1$) model is still better than others in terms of fitting $\bar{\hat{\eta}}_t$.
\begin{table}
    \centering
    \begin{tabular}{llll}
        \toprule
        & GARCH($1,1$) & GARCH($2,1$) & GARCH($2,2$) \\
        \midrule
        AIC & 3.733134 & 3.731101 & 3.738661 \\
        \bottomrule
    \end{tabular}
    \caption{AIC for GARCH($1,1$), GARCH($2,1$), and GARCH($2,2$)}
    \label{tab:table_47}
\end{table}

\begin{figure}
     \centering
     \begin{subfigure}[b]{0.3\textwidth}
         \centering
         \includegraphics[width=\textwidth]{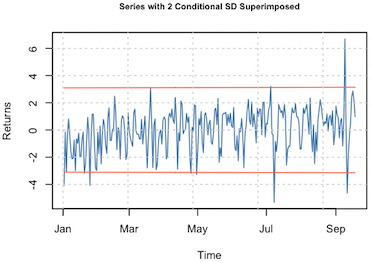}
         \caption{}
         \label{fig:essay_fig_416_1}
     \end{subfigure}
     \begin{subfigure}[b]{0.3\textwidth}
         \centering
         \includegraphics[width=\textwidth]{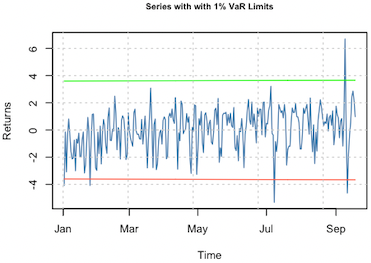}
         \caption{}
         \label{fig:essay_fig_416_2}
     \end{subfigure}
     \begin{subfigure}[b]{0.3\textwidth}
         \centering
         \includegraphics[width=\textwidth]{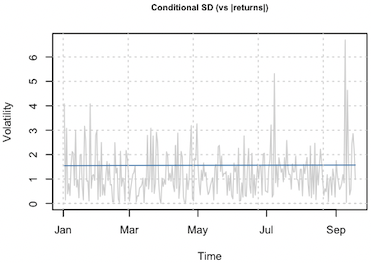}
         \caption{}
         \label{fig:essay_fig_416_3}
     \end{subfigure}     
     \begin{subfigure}[b]{0.3\textwidth}
         \centering
         \includegraphics[width=\textwidth]{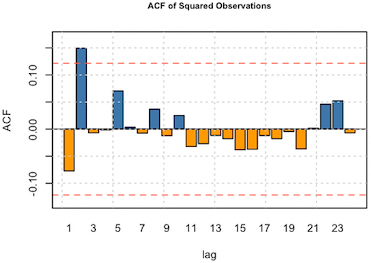}
         \caption{}
         \label{fig:essay_fig_416_4}
     \end{subfigure}
     \begin{subfigure}[b]{0.3\textwidth}
         \centering
         \includegraphics[width=\textwidth]{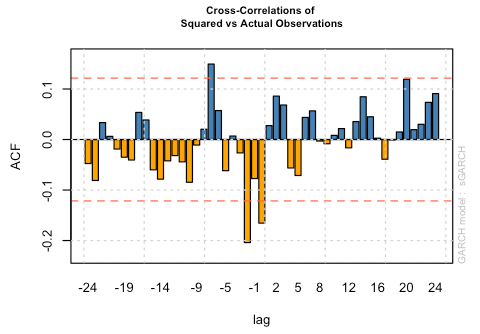}
         \caption{}
         \label{fig:essay_fig_416_5}
     \end{subfigure} 
     \begin{subfigure}[b]{0.3\textwidth}
         \centering
         \includegraphics[width=\textwidth]{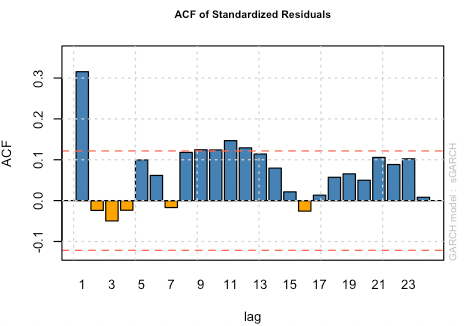}
         \caption{}
         \label{fig:essay_fig_416_6}
     \end{subfigure}     
    \caption{(a) and (b) show the residuals $\bar{\hat{\eta}}_t$ against conditional variance. (c) plots conditional standard deviation against residuals $\bar{\hat{\eta}}_t$. (d) ACF of squared $\bar{\hat{\eta}}_t$. (e) Cross-correlation of $\bar{\hat{\eta}}_t$ and squared $\bar{\hat{\eta}}_t$. (f) ACF of standardized residuals $\hat{\varepsilon}_t$}
    \label{fig:essay_fig_416}
\end{figure}

When squared residuals $\bar{\hat{\eta}}_t$ and conditional variance $\hat{\sigma}_{t \mid t-1}$ are plotted together, Figure \ref{fig:essay_fig_416_1} and \ref{fig:essay_fig_416_2}, it indicates that the model fitting fails a test for normality. The GARCH($2,1$) model cannot capture any volatility. The ACF plot of squared $\bar{\hat{\eta}}_t$, Figure \ref{fig:essay_fig_416_4}, only shows non-null autocorrelations at lag 2. Figure \ref{fig:essay_fig_416_6}, the plot of standardized residuals $\hat{\varepsilon}_t$ illustrates non-null autocorrelations at low lags, so it does not follow a normal distribution.

It is not necessary to compare the estimator $\hat{\varphi}_i$ (or $\hat{\varphi}_{1i}$, $\hat{\varphi}_{2i}$) to $\hat{\varphi}_{i\_old}$. This is because the squared residual $\bar{\hat{\eta}}_t$ does not have much high volatility, and fitting the GARCH model is not a suitable modelling method. Instead, most fMRI series are identified as AR models with orders 1, 2 and 3. It implies that the subject CC10045 can be modelled with multiple AR($\hat{u}_i$) models.

\section{Conclusions}
This paper evaluates a novel aspect of adapting an approach to handle fMRI time series data CC10045 and CC10056 for infants because infant data have “innovations” (sudden movements) associated with them, which results in jumps in the time series data. The proposed approach tries to model these jumps as shared volatility clustering across the time series (brain regions). The shared volatility clustering is modelled as GARCH models configured with the normal distribution, and the superior GARCH model is chosen based on which model achieves the smallest AIC. Each series is modelled as independent AR models. The estimations of model parameters are examined by the standard error, MSE and graphical examinations. The performance of modelling is evaluated by the performance measures MSE.

All GARCH models in this project are configured with the normal distribution. The simulation work proves that multiple dependent AR($1$) + GARCH($1,1$) time series data have shared volatility modelled successfully. The Q-Q plot of the standardized residuals shows that our prespecified distribution assumption ($\sim N(0,1)$) was correct. The graphical examination of conditional variance shows that GARCH model tends to estimate the shared volatility clustering. The parameters of AR parts are underestimated when shared volatility clustering is not weighted. After calculating the weighted shared volatility clustering by AR coefficients and averaging coefficient estimators, the estimation of AR parameters has increased accuracy. Equivalently, the MSE of parameter estimators are heavily reduced. The real-world fMRI data CC10056 with many movements is identified as multiple AR models with different orders in the range $[1, 5]$ and a GARCH($1,1$) model. The GARCH($1,1$) model tries to capture the shared volatility clustering. Also, the graphical examination of AR coefficients shows that they are estimated accurately. The real-world fMRI data CC10045 is identified as multiple AR models. However, the share volatility clustering cannot be modelled by a GARCH model. It does not reveal conditional heteroscedasticity, as it does not pass statistical tests. Moreover, the best model, GARCH($2,1$) chosen, cannot capture shared volatility clustering.

The results, therefore, imply that the weighted shared volatility improves the modelling performance of the GARCH models when handling fMRI time series data for infants with many movements across brain regions. Furthermore, averaging the AR coefficient estimators is a vital characteristic to account for when modelling multiple dependent time series.

\bibliographystyle{unsrtnat}
\bibliography{references}  

\begin{thebibliography}{18}
\providecommand{\natexlab}[1]{#1}
\providecommand{\url}[1]{\texttt{#1}}
\expandafter\ifx\csname urlstyle\endcsname\relax
  \providecommand{\doi}[1]{doi: #1}\else
  \providecommand{\doi}{doi: \begingroup \urlstyle{rm}\Url}\fi

\bibitem[Moran(2018)]{Moran:2018}
P.~A. Moran.
\newblock {Hypothesis Testing in Time Series Analysis}.
\newblock \emph{Royal Statistical Society. Journal. Series A: General}, 114\penalty0 (4):\penalty0 579--579, 12 2018.
\newblock ISSN 0035-9238.
\newblock \doi{10.2307/2981095}.
\newblock URL \url{https://doi.org/10.2307/2981095}.

\bibitem[Corbyn(2011)]{Corbyn:2011}
Judith Corbyn.
\newblock {Time Series Analysis with Applications in R, 2nd edn}.
\newblock \emph{Journal of the Royal Statistical Society Series A: Statistics in Society}, 174\penalty0 (2):\penalty0 507--507, 03 2011.
\newblock ISSN 0964-1998.
\newblock \doi{10.1111/j.1467-985X.2010.00681_4.x}.
\newblock URL \url{https://doi.org/10.1111/j.1467-985X.2010.00681\_4.x}.

\bibitem[Bollerslev(1986)]{Tim:1986}
Tim Bollerslev.
\newblock Generalized autoregressive conditional heteroskedasticity.
\newblock \emph{Journal of Econometrics}, 31\penalty0 (3):\penalty0 307--327, 1986.
\newblock ISSN 0304-4076.
\newblock \doi{https://doi.org/10.1016/0304-4076(86)90063-1}.
\newblock URL \url{https://www.sciencedirect.com/science/article/pii/0304407686900631}.

\bibitem[Zolfaghari and Gholami(2021)]{Zolfaghari:2021}
Mehdi Zolfaghari and Samad Gholami.
\newblock A hybrid approach of adaptive wavelet transform, long short-term memory and arima-garch family models for the stock index prediction.
\newblock \emph{Expert Systems with Applications}, 182:\penalty0 115149, 2021.
\newblock ISSN 0957-4174.
\newblock \doi{https://doi.org/10.1016/j.eswa.2021.115149}.
\newblock URL \url{https://www.sciencedirect.com/science/article/pii/S095741742100590X}.

\bibitem[Lin and Huang(2021)]{Lin:2021}
Xianfu Lin and Yuzhang Huang.
\newblock {Short‐Term} {High-Speed} traffic flow prediction based on {ARIMA-GARCH-M} model.
\newblock \emph{Wireless Personal Communications}, 117\penalty0 (4):\penalty0 3421--3430, April 2021.

\bibitem[Ito et~al.(2020)Ito, Hearne, and Cole]{Ito:2020}
Takuya Ito, Luke~J. Hearne, and Michael~W. Cole.
\newblock A cortical hierarchy of localized and distributed processes revealed via dissociation of task activations, connectivity changes, and intrinsic timescales.
\newblock \emph{NeuroImage}, 221:\penalty0 117141, 2020.
\newblock ISSN 1053-8119.
\newblock \doi{https://doi.org/10.1016/j.neuroimage.2020.117141}.
\newblock URL \url{https://www.sciencedirect.com/science/article/pii/S1053811920306273}.

\bibitem[Purdon et~al.(2001)Purdon, Solo, Weisskoff, and Brown]{Purdon:2001}
Patrick~L. Purdon, Victor Solo, Robert~M. Weisskoff, and Emery~N. Brown.
\newblock Locally regularized spatiotemporal modeling and model comparison for functional mri.
\newblock \emph{NeuroImage}, 14\penalty0 (4):\penalty0 912--923, 2001.
\newblock ISSN 1053-8119.
\newblock \doi{https://doi.org/10.1006/nimg.2001.0870}.
\newblock URL \url{https://www.sciencedirect.com/science/article/pii/S1053811901908705}.

\bibitem[Grady and Garrett(2014)]{Grady:2014}
Cheryl~L Grady and Douglas~D Garrett.
\newblock Understanding variability in the {BOLD} signal and why it matters for aging.
\newblock \emph{Brain Imaging Behav}, 8\penalty0 (2):\penalty0 274--283, June 2014.

\bibitem[Logothetis et~al.(2001)Logothetis, Pauls, Augath, Trinath, and Oeltermann]{Logothetis:2001}
Nikos~K Logothetis, Jon Pauls, Mark Augath, Torsten Trinath, and Axel Oeltermann.
\newblock Neurophysiological investigation of the basis of the {fMRI} signal.
\newblock \emph{Nature}, 412\penalty0 (6843):\penalty0 150--157, July 2001.

\bibitem[Logothetis(2002)]{Logothetis:2002}
Nikos~K Logothetis.
\newblock The neural basis of the blood-oxygen-level-dependent functional magnetic resonance imaging signal.
\newblock \emph{Philos Trans R Soc Lond B Biol Sci}, 357\penalty0 (1424):\penalty0 1003--1037, August 2002.

\bibitem[Kanwisher et~al.(1997)Kanwisher, McDermott, and Chun]{Kanwisher:1997}
N~Kanwisher, J~McDermott, and M~M Chun.
\newblock The fusiform face area: a module in human extrastriate cortex specialized for face perception.
\newblock \emph{J Neurosci}, 17\penalty0 (11):\penalty0 4302--4311, June 1997.

\bibitem[Haxby et~al.(2001)Haxby, Gobbini, Furey, Ishai, Schouten, and Pietrini]{Haxby:2001}
J~V Haxby, M~I Gobbini, M~L Furey, A~Ishai, J~L Schouten, and P~Pietrini.
\newblock Distributed and overlapping representations of faces and objects in ventral temporal cortex.
\newblock \emph{Science}, 293\penalty0 (5539):\penalty0 2425--2430, September 2001.

\bibitem[{Van Dijk} et~al.(2012){Van Dijk}, Sabuncu, and Buckner]{VanDijk:2012}
Koene~R.A. {Van Dijk}, Mert~R. Sabuncu, and Randy~L. Buckner.
\newblock The influence of head motion on intrinsic functional connectivity mri.
\newblock \emph{NeuroImage}, 59\penalty0 (1):\penalty0 431--438, 2012.
\newblock ISSN 1053-8119.
\newblock \doi{https://doi.org/10.1016/j.neuroimage.2011.07.044}.
\newblock URL \url{https://www.sciencedirect.com/science/article/pii/S1053811911008214}.
\newblock Neuroergonomics: The human brain in action and at work.

\bibitem[McLeod and Li(1983)]{McLeod:1983}
A.~I. McLeod and W.~K. Li.
\newblock Diagnostic checking arma time series models using squared-residual autocorrelations.
\newblock \emph{Journal of Time Series Analysis}, 4\penalty0 (4):\penalty0 269--273, 1983.
\newblock \doi{https://doi.org/10.1111/j.1467-9892.1983.tb00373.x}.
\newblock URL \url{https://onlinelibrary.wiley.com/doi/abs/10.1111/j.1467-9892.1983.tb00373.x}.

\bibitem[Engle(1982)]{Engle:1982}
Robert~F. Engle.
\newblock Autoregressive conditional heteroscedasticity with estimates of the variance of united kingdom inflation.
\newblock \emph{Econometrica}, 50\penalty0 (4):\penalty0 987--1007, 1982.
\newblock ISSN 00129682, 14680262.
\newblock URL \url{http://www.jstor.org/stable/1912773}.

\bibitem[LJUNG and BOX(1978)]{LJUNG:1978}
G.~M. LJUNG and G.~E.~P. BOX.
\newblock {On a measure of lack of fit in time series models}.
\newblock \emph{Biometrika}, 65\penalty0 (2):\penalty0 297--303, 08 1978.
\newblock ISSN 0006-3444.
\newblock \doi{10.1093/biomet/65.2.297}.
\newblock URL \url{https://doi.org/10.1093/biomet/65.2.297}.

\bibitem[Naik et~al.(2020)Naik, Mohan, and Jha]{Naik:2020}
Nagaraj Naik, Biju~R Mohan, and Rajat~Aayush Jha.
\newblock Garch-model identification based on performance of information criteria.
\newblock \emph{Procedia Computer Science}, 171:\penalty0 1935--1942, 2020.
\newblock ISSN 1877-0509.
\newblock \doi{https://doi.org/10.1016/j.procs.2020.04.207}.
\newblock URL \url{https://www.sciencedirect.com/science/article/pii/S1877050920311893}.
\newblock Third International Conference on Computing and Network Communications (CoCoNet'19).

\bibitem[WILK and GNANADESIKAN(1968)]{Wilk:1968}
M.~B. WILK and R.~GNANADESIKAN.
\newblock {Probability plotting methods for the analysis for the analysis of data}.
\newblock \emph{Biometrika}, 55\penalty0 (1):\penalty0 1--17, 03 1968.
\newblock ISSN 0006-3444.
\newblock \doi{10.1093/biomet/55.1.1}.
\newblock URL \url{https://doi.org/10.1093/biomet/55.1.1}.

\end{thebibliography}






\end{document}